\documentclass[12pt]{article}
 \usepackage{amsmath, amssymb, amsthm, float, fullpage, graphicx, parskip, subcaption}
\usepackage{xcolor}
\usepackage{authblk}

\usepackage{hyperref}

\definecolor{PennRed}{RGB}{152, 30 50}
\definecolor{PennBlue}{RGB}{0, 44, 119}
\definecolor{PennGreen}{RGB}{94, 179,70}
\definecolor{PennViolet}{RGB}{141, 76, 145}
\definecolor{PennSkyBlue}{RGB}{14, 118, 188}
\definecolor{PennOrange}{RGB}{243, 117, 58}
\definecolor{PennBrightRed}{RGB}{223,82, 78}

\hypersetup{pdfborder = {0 0 0.5 [3 3]}, colorlinks = true, linkcolor = PennBrightRed, citecolor = PennSkyBlue}

\title{Estimating an NBA player's impact on his team's chances of winning}
\author[]{Sameer K. Deshpande}
\author[]{Shane T. Jensen}
\affil[]{The Wharton School \\ University of Pennsylvania}
\date{}
\begin{document}
\maketitle

\section{Introduction}
\label{sec:introduction}
Determining which National Basketball Association (NBA) players do the most to help their teams win games is perhaps the most natural question in basketball analytics.
Traditionally, one quantifies the notion of helping teams win with a scoring statistic like points-per-game or true shooting percentage, a function of point differential like Adjusted Plus-Minus (see, e.g., \cite{Rosenbaum2004}, \cite{IlardiBarzilai2008}) and variants thereof, or some combination of box score statistics and pace of play like the Player Efficiency Rating (PER) of \cite{Hollinger2004}.

While these metrics are informative, we observe that they ignore the \textit{context} in which players perform.
As a result, they can artificially inflate the importance of performance in low-leverage situations, when the outcome of the game is essentially decided, while simultaneously deflating the importance of high-leverage performance, when the final outcome is still in question.
For instance, point differential-based metrics model the home team's lead dropping from 5 points to 0 points in the last minute of the first half in exactly the same way that they model the home team's lead dropping from 30 points to 25 points in the last minute of the second half.
In both of these scenarios, the home team's point differential is -5 points but, as we will see in Section~\ref{sec:win_probability}, the home team's chance of winning the game dropped from 72\% to 56\% in the first scenario while it remained constant at 100\% in the second. 
We argue that a player's performance in the second scenario has no impact on the final outcome and should therefore not be treated comparably to performance in the first.
We address this issue by proposing a win probability framework and linear regression model to estimate each player's contribution to his team's overall chance of winning games.

The use of win probability to evaluate the performance of professional athletes dates back at least to \cite{MillsMills1970}, who evaluated Major League Baseball players.
As \cite{Studeman2004} observes, their Player Wins Average methodology has been re-introduced several times since, most notably as Win Probability Added (WPA).
To compute WPA, one starts with an estimate of a team's probability of winning the game at various game states.
For each plate appearance, one then credits the pitcher and batter with the resulting change in their respective team's win probability and then sums these contributions over the course of a season to determine how involved a player was in his team's wins (or losses).
A natural extension of the WPA methodology to basketball would be to measure the change in the win probability from the time a player enters the game to the time he is substituted out of the game and then sum these changes over the course of a season.
Such an extension is identical to the traditional plus-minus statistic except that it is computed on the scale of win probability instead of points scored.

An inherent weakness of using plus-minus (on both the point and win probability scales) to assess a player's performance is that a player's plus-minus statistic necessarily depends on the contributions of his teammates and opponents.
According to \cite{GramacyJensenTaddy2013}, since the the quality of any individual player's pool of teammates and opponents can vary dramatically, ``the marginal plus-minus for individual players are inherently polluted.''
To overcome this difficulty, \cite{Rosenbaum2004} introduced Adjusted Plus-Minus to estimate the average number of points a player scores per 100 possession after controlling for his opponents and teammates.
To compute Adjusted Plus-Minus, one first breaks the game into several ``shifts,'' periods of play between substitutions, and measures both the point differential and total number of possessions in each shift.
One then regresses the point differential per 100 possessions from the shift onto indicators corresponding to the ten players on the court.

We propose instead to regress the change in the home team's win probability during a shift onto signed indicators corresponding to the five home team players and five away team players in order to estimate each player's \textit{partial effect} on his team's chances of winning.
Briefly, if we denote the change in the home team's win probability in the $i^{th}$ shift by $y_{i},$ we then model the expected change in win probability given the players on the court, as
\begin{equation}
\label{eq:general_model}
E[y_{i} | \mathbf{h}_{i}, \mathbf{a}_{i}] = \mu_{i} + \theta_{h_{i1}} + \cdots + \theta_{h_{i5}} - \theta_{a_{i1}} - \cdots - \theta_{a_{i5}}
\end{equation}
where $\theta = \left( \theta_{1}, \ldots, \theta_{488}\right)^{\top}$ is the vector of player partial effects and  $\mathbf{h}_{i} = \left\{ h_{i1}, \ldots, h_{i5} \right\}$ and $\mathbf{a}_{i} = \left\{ a_{i1}, \ldots, a_{i5} \right\}$ are indices on $\theta$ corresponding to the home team (h) and away team (a) players.
The intercept term $\mu_{i}$ may depend on additional covariates, such as team indicators.

Fitting the model in Equation~\ref{eq:general_model} is complicated by the fact that we have a relatively large number of covariates (viz. a total of 488 players in the 2013-2014 season) displaying a high degree of collinearity, since some players are frequently on the court together.
This can lead to imprecise estimates of player partial effects with very large standard errors.
Regularization, which shrinks the estimates of the components of $\theta$ towards zero, is therefore necessary to promote numerical stability for each partial effect estimate.

We take a Bayesian approach, which involves specifying a \textit{prior} distribution with mode at zero on each partial effect and using play-by-play data from the 2013-14 season to update these priors to get a \textit{posterior} distribution of the partial effects.
As \cite{KyungGillGhoshCasella2010} argue, the Bayesian formulation of regularized regression produces valid and tractable standard errors, unlike popular frequentist techniques like the lasso of \cite{Tibshirani1996}.
This enables us to quantify the joint uncertainty of our partial effect estimates in a natural fashion.

Our proposed methodology produces a \textit{retrospective} measure of individual player contributions and does not attempt to measure a player's latent ability or talent.
Our estimates of player partial effect are context-dependent, making them unsuitable for forecasting future performance since the context in which a player plays can vary season-to-season and even week-to-week.
Nevertheless, because our proposal is context-dependent, we feel that it provides a more appropriate accounting of what actually happened than existing player-evaluation metrics like PER and ESPN's Real Plus/Minus (RPM). 
Taken together with such existing metrics, our estimates of player effect can provide insight into whether coaches are dividing playing time most effectively and help understand the extent to which a player's individual performance translate to wins for his team.

The rest of this paper is organized as follows.
We detail our data and regression model in Section~\ref{sec:data_model_methods} and describe our estimation of win probability in Section~\ref{sec:win_probability}.
Section~\ref{sec:full_posterior_analysis} presents a full Bayesian analysis of the joint uncertainty about player effects.
In Section~\ref{sec:comparing_players} we introduce \textit{leverage profiles} to measure the similarity between the contexts in which two players performed. 
These profiles enable us to make meaningful comparisons of players based on their partial effect estimates. 
In keeping with the examples of other player evaluation metrics, we propose two rank-orderings of players using their partial effects.
In Section~\ref{sec:impact_ranking}, we rank players on a team-by-team basis, allowing us to determine each player's relative value to his team.
In Section~\ref{sec:impact_score}, we present a single ranking of all players which balances a player's partial effect against the posterior uncertainty in estimating his effect.
We extend our analysis of player partial effects in Section~\ref{sec:lineup_comparison} to five-man lineups and consider how various lineups matchup against each other.
We conclude in Section~\ref{sec:discussion} with a discussion of our results and several extensions.

\section{Data, Models, and Methods}
\label{sec:data_model_methods}
Like Adjusted Plus/Minus, we break each game into shifts: periods of play between successive substitutions when the ten players on the court is unchanged.
During the 2013--2014 regular season, a typical game consisted of around 31 shifts. 
In order to determine which players are on the court during each shift, we use play-by-play data obtained from ESPN
for 8365 of the 9840 (85\%) of the scheduled regular season games in each of the eight seasons between 2006 and 2014.
The play-by-play data for the remaining 15\% of games were either incomplete or missing altogether.
The majority of the missing games were from the first half of the time window considered.
To the best of our knowledge, our dataset does not systematically exclude games from certain teams or certain types of games (early-season vs late-season, close game vs blow-out).
Using the data from the 2006--2007 season to 2012--2013 season, we estimate the home team's win probability as a function of its lead and the time elapsed. 
With these win probability estimates, we then compute the change in the home team's win probability during each of $n = 35,799$ shifts in the 2013--2014 regular season.
We denote the change in the home team's win probability during the $i^{th}$ shift by $y_{i}.$
This change in win probability can be directly attributed to the performance of the 10 players on the court during that shift. 
Thus, to measure each individual player's impact on the change in win probability, we regress $y_{i}$ onto indicator variables corresponding to which of the 488 players were on the court during the $i^{th}$ shift. 
We model
\begin{equation}
\label{eq:player_team_model}
y_{i}|\mathbf{h_{i}, a_{i}} = \mu + \theta_{h_{i1}} + \cdots + \theta_{h_{i5}} - \theta_{a_{i1}} - \cdots \theta_{a_{i5}} + \tau_{H_{i}} - \tau_{A_{i}} + \sigma\varepsilon_{i},
\end{equation}
where $\theta = \left( \theta_{1}, \ldots, \theta_{488}\right)^{\top}$ is the vector of partial effects for the 488 players, $\tau = \left(\tau_{1}, \ldots, \tau_{30}\right)^{\top}$ is a vector of partial effects for the 30 teams, with $\mathbf{h}_{i} = \left\{ h_{i1}, \ldots, h_{i5} \right\}$ and $\mathbf{a}_{i} = \left\{ a_{i1}, \ldots, a_{i5} \right\}$ are indices on $\theta$ corresponding to the home team (h) and away team (a) players, $H_{i}$  and $A_{i}$ are indices  on $\tau$ corresponding to which teams are playing in shift $i$, and the $\varepsilon_{i}$ are independent standard normal random variables.
We view $\mu$ as a league-average ``home-court advantage'' and $\sigma$ as a measure of the variability in $y_{i}$ that arises from both the uncertainty in measuring $y_{i}$ and the inherent variability in win probability that cannot be explained by the performance of the players on the court.
Since we are including team effects in Equation~\ref{eq:player_team_model}, each player's partial effect is measured relative to his team's average, so that players are not overly penalized.

\subsection{Estimation of Win Probability}
\label{sec:win_probability}
\begin{figure}[!h]
\centering
\includegraphics{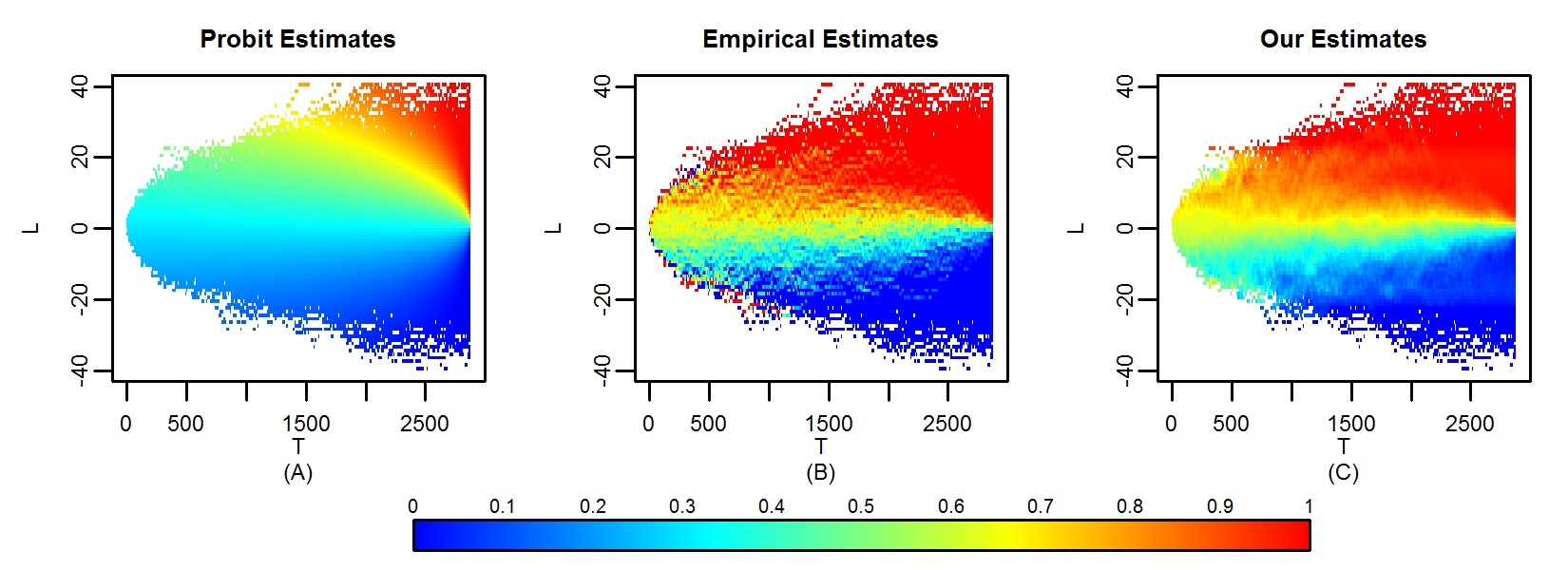}
\caption{Various estimates of $p_{T,L}$. The probit estimates in (A), while smooth, do not agree with the empirical win probabilities shown in (B). 
Our estimates, shown in (C), are closer in value to the empirical estimates than are those in (A) but are much smoother than the empirical estimates.}
\label{fig:winProb}
\end{figure}

In order to fit such a regression model, we must begin with an estimate of the probability that the home team wins the game after leading by $L$ points after $T$ seconds, which we denote by $p_{T,L}.$
Estimating win probability at specific intermediate times during a game is not a new problem; indeed, \cite{Lindsey1963} estimated win probabilities in baseball in the 1960's and \cite{Stern1994} introduced a probit regression model to estimate $p_{T,L}.$
\cite{MayminMayminShen2012} expanded on that probit model to study when to take starters in foul trouble out of a game, \cite{Bashuk2012} considered empirical estimates of win probability to predict team performance in college basketball, and \cite{Pettigrew2015} recently introduced a parametric model to estimate win probability in hockey.
Intuitively, we believe that $p_{T,L}$ is a smooth function of both $T$ and $L$; for a fixed lead, the win probability should be relatively constant for a small duration of time.
By construction, the probit model of \cite{Stern1994} produces a smooth estimate of the win probability and the estimates based on all games from the 2006--07 to 2012--13 regular seasons are shown in Figure~\ref{fig:winProb}(A), where the color of the unit cell $[T,T+1] \times [L, L + 1]$ corresponds to the estimated value of $p_{T,L}.$

To get a sense of how well the probit estimates fit the observed data, we can compare them to the empirical estimates of $p_{T,L}$ given by the proportion of times that the home team has won after leading by $L$ points after $T$ seconds. 
The empirical estimates of $p_{T,L}$ are shown in Figure~\ref{fig:winProb}(B). 

We see immediately that the empirical estimates are rather different than the probit estimates: for positive $L$, the probit estimate of $p_{T,L}$ tends to be much smaller than the empirical estimate of $p_{T,L}$ and for negative L, the probit estimates tend to overestimate $p_{T,L}$. 
This discrepancy arises primarily because the probit model is fit using only data from the ends of the first three quarters and does not incorporate any other intermediate times. 
Additionally, the probit model imposes several rather strong assumptions about the evolution of the win probability as the game progresses. 
As a result, we find the empirical estimates much more compelling than the probit estimates. 
Despite this, we observe in Figure~\ref{fig:winProb}(B) that the empirical estimates are much less smooth than the probit estimates.
Also worrying are the extreme and incongruous estimates near the edges of the colored region in Figure~\ref{fig:winProb}(B). 
For instance, the empirical estimates suggest that the home team will always win the game if they trailed by 18 points after five minutes of play. 
Upon further inspection, we find that the home team trailed by 18 points after five minutes exactly once in the seven season span from 2006 to 2013 and they happened to win that game. 
In other words, the empirical estimates are rather sensitive to small sample size leading to extreme values which can heavily bias our response variables $y_{i}$ in Equation~\ref{eq:player_team_model}.

To address these small sample issues in the empirical estimate, we propose a middle ground between the empirical and probit estimates. 
In particular, we let $N_{T,L}$ be the number of games in which the home team has led by $\ell$ points after $t$ seconds where $T-h_{t} \leq t \leq T+h_{t}$ and $L-h_{l} \leq \ell \leq L+h_{l},$ where $h_{t}$ and $h_{l}$ are positive integers.
We then let $n_{T,L}$ be the number of games which the home team won in this window and model $n_{T,L}$ as a Binomial$(N_{T,L}, p_{T,L})$ random variable.
This modeling approach is based on the assumption that the win probability is relatively constant over a small window in the $(T,L)$-plane.
The choice of $h_{t}$ and $h_{l}$ dictate how many game states worth of information is used to estimate $p_{T,L}$ and larger choices of both will yield, in general, smoother estimates of $p_{T,L}.$
Since very few offensive possession last six seconds or less and since no offensive possession can result in more than four points, we argue that the win probability should be relatively constant in the window $[T - 3, T + 3] \times [L - 2, L + 2]$ and we take $h_{t} = 3, h_{l} = 2.$

We place a conjugate Beta($\alpha_{T,L}, \beta_{T,L})$ prior on $p_{T,L}$ and estimate $p_{T,L}$ with the resulting posterior mean $\hat{p}_{T,L},$ given by
$$
\hat{p}_{T,L} = \frac{n_{T,L} + \alpha_{T,L}}{N_{T,L} + \alpha_{T,L} + \beta_{T,L}}.
$$
The value of $y_{i}$ in Equation~\ref{eq:player_team_model} is the difference between the estimated win probability at the end of the shift and at the start of the shift. 

Based on the above expression, we can interpret $\alpha_{T,L}$ and $\beta_{T,L}$ as ``pseudo-wins'' and ``pseudo-losses'' added to the observed counts of home team wins and losses in the window $[T - 3, T + 3] \times [L - 2, L + 2].$
The addition of these ``pseudo-games'' tends to shrink the original empirical estimates of $p_{T,L}$ towards $\frac{\alpha_{T,L}}{\alpha_{T,L} + \beta_{T,L}}.$
To specify $\alpha_{T,L}$ and $\beta_{T,L},$ it is enough to describe how many pseudo-wins and pseudo-losses we add to each of the 35 unit cells $[t, +1] \times [\ell, \ell + 1]$ in the window $[T - 3, T + 3] \times [L - 2, L + 2].$
We add a total of 10 pseudo-games to each unit cell, but the specific number of pseudo-wins depends on the value of $\ell.$
For $\ell < -20,$ we add 10 pseudo-losses and no pseudo-wins and for $\ell > 20,$ we add 10 pseudo-wins and no pseudo-losses.
For the remaining values of $\ell,$ we add five pseudo-wins and five pseudo-losses.
Since we add 10 pseudo-games to each cell, we add a total of $\alpha_{T,L} + \beta_{T,L} = 350$ pseudo-games the window $[T - 3, T + 3] \times [L - 2, L + 2].$
We note that this procedure does not ensure that our estimated win probabilities are monotonic in lead and time.
However, the empirical win probabilities are far from monotonic themselves, and our procedure does mitigate many of these departures by smoothing over the window $[T- 3, T + 3] \times [L - 2, L + 2].$

We find that for most combinations of $T$ and $L$, $N_{T,L}$ is much greater than 350; for instance, at $T = 423,$ we observe $N_{T,L} =$ 4018, 11375, 17724, 14588, and 5460 for $L =$ -10, -5, 0, 5, and 10, respectively.
In these cases, the value of $\hat{p}_{T,L}$ is driven more by the observed data than by the values of $\alpha_{T,L}$ and $\beta_{T,L}.$
Moreover, in such cases, the uncertainty of our estimate $\hat{p}_{T,L}$, which can be measured by the posterior standard deviation of $p_{T,L}$ is exceeding small: for $T = 423$ and $-10 \leq L \leq 10,$ the posterior standard deviation of $p_{T,L}$, is between 0.003 and 0.007.
When $N_{T,L}$ is comparable to or much smaller than 350, the values of $\alpha_{T,L}$ and $\beta_{T,L}$ exert more influence on the value of $\hat{p}_{T,L}.$
The increased influence of the prior on $\hat{p}_{T,L}$ in such rare game states helps smooth over the extreme discontinuities that are present in the empirical win probability estimates above.
In these situations, there is a larger degree of uncertainty in our estimate of $\hat{p}_{T,L},$ but we find that the posterior standard deviation of $p_{T,L}$ never exceeds 0.035. 
The uncertainty in our estimation of $p_{T,L}$ leads to additional uncertainty in the $y_{i}$'s, akin to measurement error.
The error term in Equation~\ref{eq:player_team_model} is meant to capture this additional uncertainty, as well as any inherent variation in the change in win probability unexplained by the players on the court.

\subsection{Bayesian Linear Regression of Player Effects}
\label{sec:bayesian_approach}

As mentioned in Section~\ref{sec:introduction}, we take a Bayesian approach to fitting the model in Equation~\ref{eq:player_team_model}.
Because we have a large number of covariates displaying a high degree of collinearity, a regularization prior that shrinks each component of $\theta$ towards zero is needed to promote stability for each partial effect.
Popular choices of regularization priors on the components $\theta_{j}$ include a normal prior, which corresponds to an $\ell_{2}$ penalty, or a Laplace prior, which corresponds to an $\ell_{1}$ penalty. 
\cite{ThomasVenturaJensenMa2013} also consider a Laplace-Gaussian prior, which combines both $\ell_{2}$ and $\ell_{1}$ penalties.
Maximum a posteriori estimation with respect to these priors correspond to ridge, lasso, and elastic net regression, respectively. 

We choose to use the Laplace prior, which was also considered by \cite{ThomasVenturaJensenMa2013} to derive rankings of National Hockey League players. 
Between the normal and Laplace prior, we choose to use the Laplace prior since it tends to pull smaller partial effects towards zero faster than the normal prior, as noted by \cite{ParkCasella2008}.
We are thus able to use the existing \texttt{R} implementation of \cite{ParkCasella2008}'s Gibbs sampler in the \texttt{monomvn} package.
Though the elastic net is better suited for regression problems in which there are groups of highly correlated predictors than is the lasso (\cite{ZouHastie2005}), there is no widely-available Gibbs sampler and the computational challenge of implementation offsets the additional benefit we gain from using the Laplace-Gaussian prior. 
We let $\mathbf{P}^{i}$ be a vector indicating which players are on the court during shift $i$ so that its  $j^{th}$ entry, $\mathbf{P}^{i}_{j},$ is equal to 1 if player $j$ is on the home team, -1 if player $j$ is on the away team, and 0 otherwise.
Similarly, we let $\mathbf{T}^{i}$ be a vector indicating which teams are playing during shift $i$ so that its $k^{th}$ entry, $\mathbf{T}^{i}_{k},$ is equal to 1 if team $k$ is the home team, -1 is team $k$ is the away team, and 0 otherwise. 
Conditional on $\mathbf{P}^{i}$ and $\mathbf{T}^{i},$ we model
$$
y_{i}|\mathbf{P}^{i}, \mathbf{T}^{i} \sim N\left(\mu + \mathbf{P}^{i^{\top}}\theta + \mathbf{T}^{i^{\top}}\tau, \sigma^{2}\right).
$$

We place independent Laplacian priors on each component of $\theta$ and $\tau$, conditional on the corresponding noise parameters $\sigma^{2}.$
The conditional prior densities of $\left(\theta, \tau \right)$ given $\sigma^{2}$ is given by
$$
p(\theta, \tau | \sigma^{2}) \propto \left[\left( \frac{\lambda}{\sigma}\right)^{488} \times \exp{\left\{-\frac{\lambda}{2\sigma}\sum_{j = 1}^{488}{|\theta_{j}|}\right\}}\right] \times \left[ \left( \frac{\lambda}{\sigma}\right)^{30} \times \exp{\left\{-\frac{\lambda}{2\sigma}\sum_{k = 1}^{30}{|\tau_{k}|}\right\}}\right],
$$
where $\lambda > 0$ is a sparsity parameter that governs how much each component of $\theta$ is shrunk towards zero.
We further place a flat prior on $\mu,$ a Gamma$(r,\delta)$ hyper-prior on $\lambda^{2},$ and non-informative hyper-priors on $\sigma^{2}, r,$ and $\delta.$

Because of the hierarchical structure of our model, the joint posterior distribution of $(\mu, \theta, \tau, \sigma^{2})$ is not analytically tractable and we must instead rely on a Markov Chain Monte Carlo (MCMC) simulation to estimate the posterior distribution.
We use the Gibbs sampler described by \cite{ParkCasella2008} that is implemented in the \texttt{monomvn} package in \texttt{R}.
We note that our prior specification is the default setting for this implementation.

\begin{figure}[!h]
\centering
\includegraphics{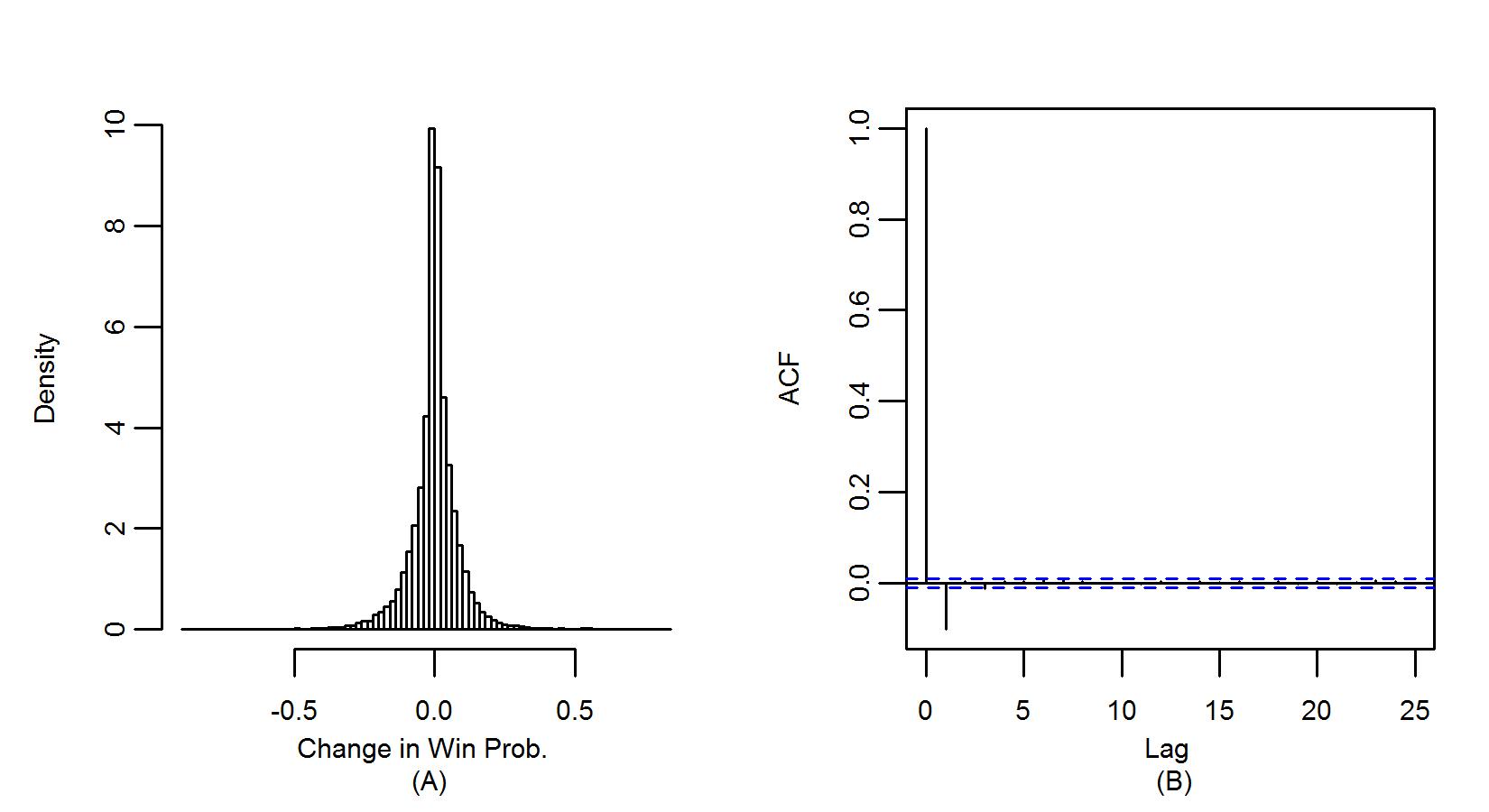}
\caption{Histogram and autocorrelation plot of the $y_{i}$'s.}
\label{fig:hist_acf_winProb}
\end{figure}

In specifying this regression model, we make several strong assumptions.
First, we assume that the $y_{i}'$s are independent. 
Since it is generally not the case that all ten players are substituted out of the game at the end of the shift, it is reasonable to expect that there will be some autocorrelation structure among the $y_{i}$'s. 
Indeed, as seen in the autocorrelation plot in Figure~\ref{fig:hist_acf_winProb}(B), we observe a small amount of autocorrelation (-0.1) between $y_{i}$ and $y_{i+1}.$
We also observe that there is no significant autocorrelation at larger lags.
While the independence assumption is not technically correct, the lack of persistent autocorrelation and the relatively weak correlation between $y_{i}$ and $y_{i+1},$ make the assumption somewhat more palatable.

Our second modeling assumption is that, conditional on $\left(\mathbf{P}_{i}, \mathbf{T}_{i}\right),$ the $y_{i}$'s are Gaussian with constant variance.
This conditional Gaussian assumption does not imply that the $y_{i}$'s are marginally Gaussian (which does not seem to be the case in Figure~\ref{fig:hist_acf_winProb}(A)).
Despite the fact that we have 35,799 shifts in our dataset, we find that there are 29,453 unique combinations of ten players on the court.
Thus, we only observe a few instances of each unique $\left(\mathbf{P}_{i}, \mathbf{T}_{i}\right)$ making it difficult to assess the conditional normality assumption directly. 
The limited number of each $\left(\mathbf{P}_{i}, \mathbf{T}_{i}\right)$ also makes it difficult to check the assumption of constance variance of the $y_{i}$'s conditional on $\left(\mathbf{P}_{i}, \mathbf{T}_{i}\right).$
In the Appendix, we explore several transformations and alternative specifications of the $y_{i}$'s, but do not find alternatives that match these assumptions better than our current specification. 

\begin{figure}[!h]
\centering
\includegraphics{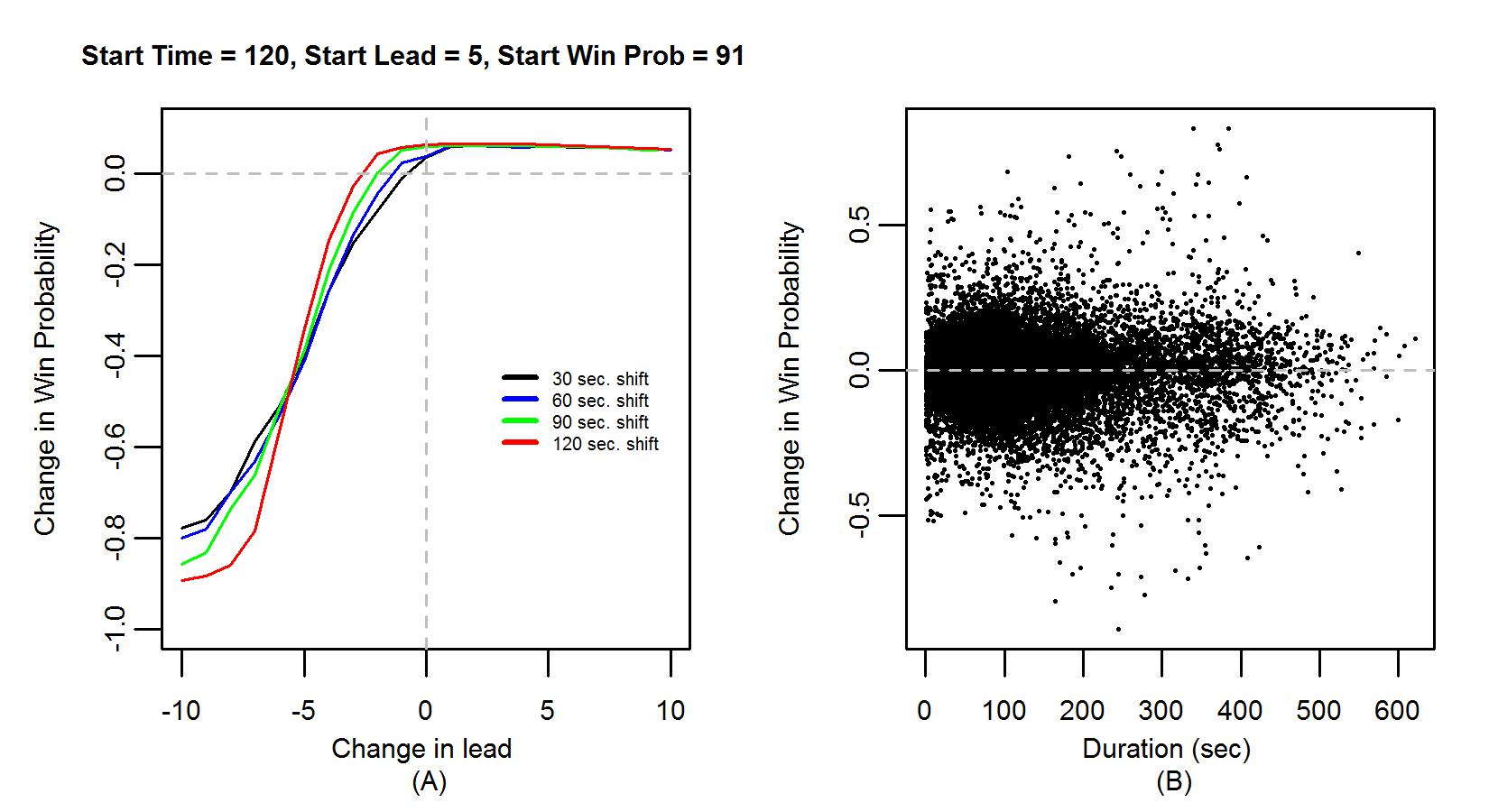}
\caption{Change in win probability plotted against shift duration}
\label{fig:duration_y}
\end{figure}

At this point, it is also worth mentioning that our model does not explicitly include the duration of each shift as a predictor, despite the fact that $y_{i}$ depends on shift length. 
Figure~\ref{fig:duration_y}(A) shows the change in win probability associated with varying shift durations and varying lead changes.
Quite clearly, we see that the curves in Figure~\ref{fig:duration_y}(A) are different, indicating a dependence between $y_{i}$ and shift duration, although we see in Figure~\ref{fig:duration_y}(B) that the overall correlation is quite small.
On a conceptual level, a player's performance in a 15 second shift during which his team's win probability increases by 20\% has the same impact on his team's chances of winning had the shift lasted 30 seconds.
Since our ultimate goal is to estimate each player's individual impact, as opposed to his playing time-adjusted impact or per-minute impact, including shift duration as an additional predictor distorts the desired interpretation of player partial effects.
In fact, we assert that the change in win probability as an outcome variable is the most natural way to account for the effect of shift duration on a player's overall impact on the court. 

\section{Full Posterior Analysis}
\label{sec:full_posterior_analysis}

\begin{figure}[!h]
\centering
\includegraphics[width = 4.5in]{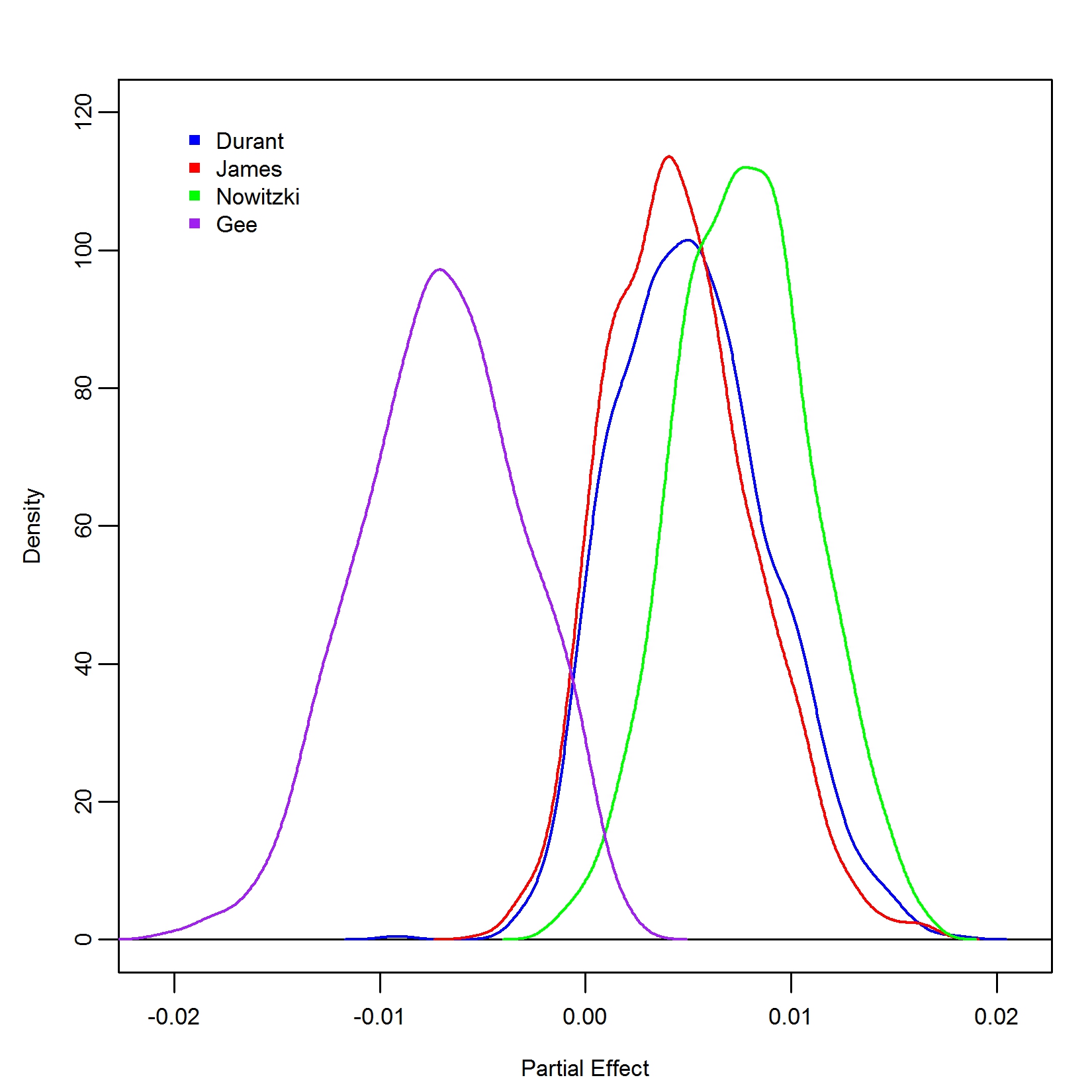}
\caption{Approximate posterior densities of several players' partial effects.}
\label{fig:player_effect_density}
\end{figure}

We use the Gibbs sampler function \texttt{lasso} in the \texttt{monomvn R} package to obtain 1000 independent samples from the full posterior distribution of $\left(\mu, \theta, \tau, \sigma^{2}\right).$
With these samples, we can approximate the marginal posterior density of each player's partial effect using a standard kernel density estimator. 
Figure~\ref{fig:player_effect_density} shows the estimated posterior densities of the partial effects of several players.

We see that these densities are almost entirely supported within the range [-0.02, 0.02], indicating that it is unlikely that any individual player, over the course of a single shift, is able to improve (or hurt) his team's chances of winning the game by more than a percentage point or two.
This is partly due to our regularization prior, which tends to pull the components of $\theta$ and $\tau$ towards zero, and to the fact that the $y_{i}$'s are tightly concentrated near zero.
Nevertheless, though our estimates of each player's partial effect are small, we still see considerable heterogeneity in the approximate posterior densities.
Most strikingly, we see that the posterior distribution of Dirk Nowitzki's partial effect is mostly supported on the positive axis (in 991 out of our 1000 posterior samples, his effect is positive) while the posterior distribution of Alonzo Gee's partial effect is mostly supported on the negative axis (his partial effect is negative in 976 out of 1000 posterior samples).

Intuitively, we can measure a player's ``value'' by his partial effect on his team's chances of winning.
Among the players in Figure~\ref{fig:player_effect_density}, we see that Nowitzki was the most valuable since his density lies further to the right than any other player's.
However, there is considerable overlap in the support of his density and that of Kevin Durant, making it difficult to determine who is decidedly the ``most valuable.''
Indeed, we find that Nowitzki's partial effect is greater than Kevin Durant's in 692 out of 1000 posterior samples.
We also observe high similarity in the posterior densities of Durant and LeBron James, who finished first and second, respectively, in voting for the 2013-14 Most Valuable Player (MVP) award.
On closer inspection, we find that Durant's partial effect is greater than James' in only 554 of the 1000 posterior samples, indicating that, by the end of the 2013-14 regular season, Durant and James had very nearly the same impact on their teams' chances of winning, with Durant enjoying a rather slight advantage.
In the context of the MVP award, then, our results would suggest that Durant is only slightly more deserving than James, but Nowitzki is more deserving than both Durant and James.

\begin{figure}[!h]
\centering
\includegraphics{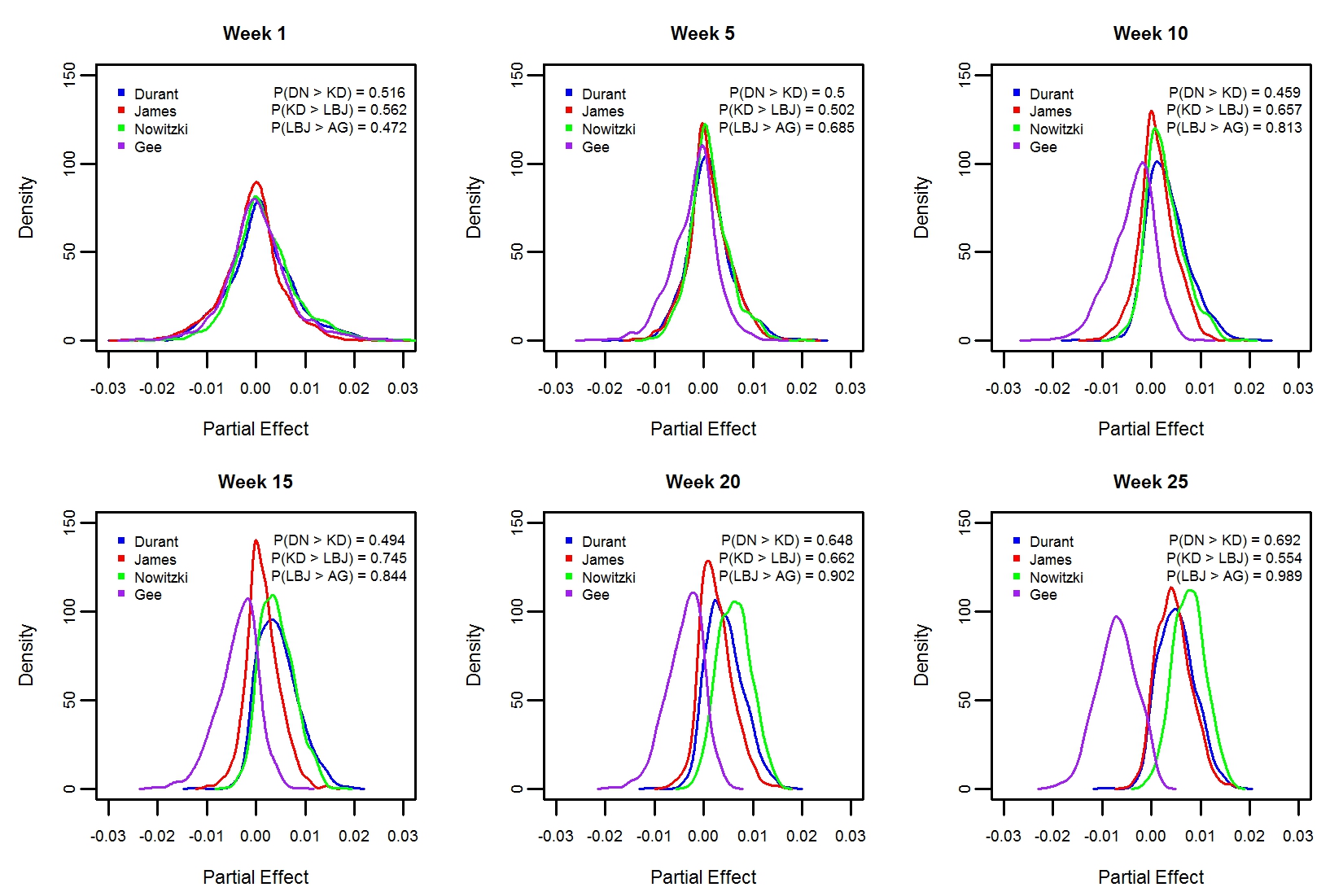}
\caption{Approximate posterior densities of Kevin Durant's, LeBron James', and Dirk Nowitzki's partial effect as the season progresses}
\label{fig:weekly_comparison}
\end{figure}

We can also track how the posterior distribution of player partial effects evolve over the course of the season, which helps to determine how many games worth of data is necessary to start differentiating the partial effects of different players.
Figure~\ref{fig:weekly_comparison} show the approximate posterior densities of Durant, Gee, James, and Nowitzki after weeks 1, 5, 10, 15, 20, and 25 of the season.

Through the first five weeks of the season, the posterior distributions of each player's partial effects are virtually identical.
However, after ten weeks, we begin to see some separation, with Gee's density moving towards the left and Durant's density moving towards the right. 
This suggests that we need at least ten weeks worth of data (approximately 30 - 35 games) in order to identify differences in player partial effects.
We see a rather considerable gap between Durant's and James' densities by week 15 and we observe that Durant's partial effect is greater than James' in nearly 75\% of the posterior samples up to that time.
Over the next 10 weeks, though, this gap shrinks considerably: visually, the two posterior densities become increasingly indistinguishable and the proportion of posterior samples in which Durant's partial effect is greater than James' shrinks back towards 0.5.
This mirrors the general consensus described by \cite{Ballentine2014} and \cite{Buckley2014} about how the race for the MVP award evolved: Durant was the clear front-runner for the MVP award by late January (approximately week 13 of the season) but many reporters declared the race much closer after James' historic performances in weeks 18 and 19 (including multiple 40-point performances and a 61-point performance against Charlotte).
We also see that the separation between Nowitzki's density and Durant's density increases between weeks 15 and 20.

\subsection{Comparing Players}
\label{sec:comparing_players}

\begin{figure}[!h]
\centering
\includegraphics[width = 4.5in]{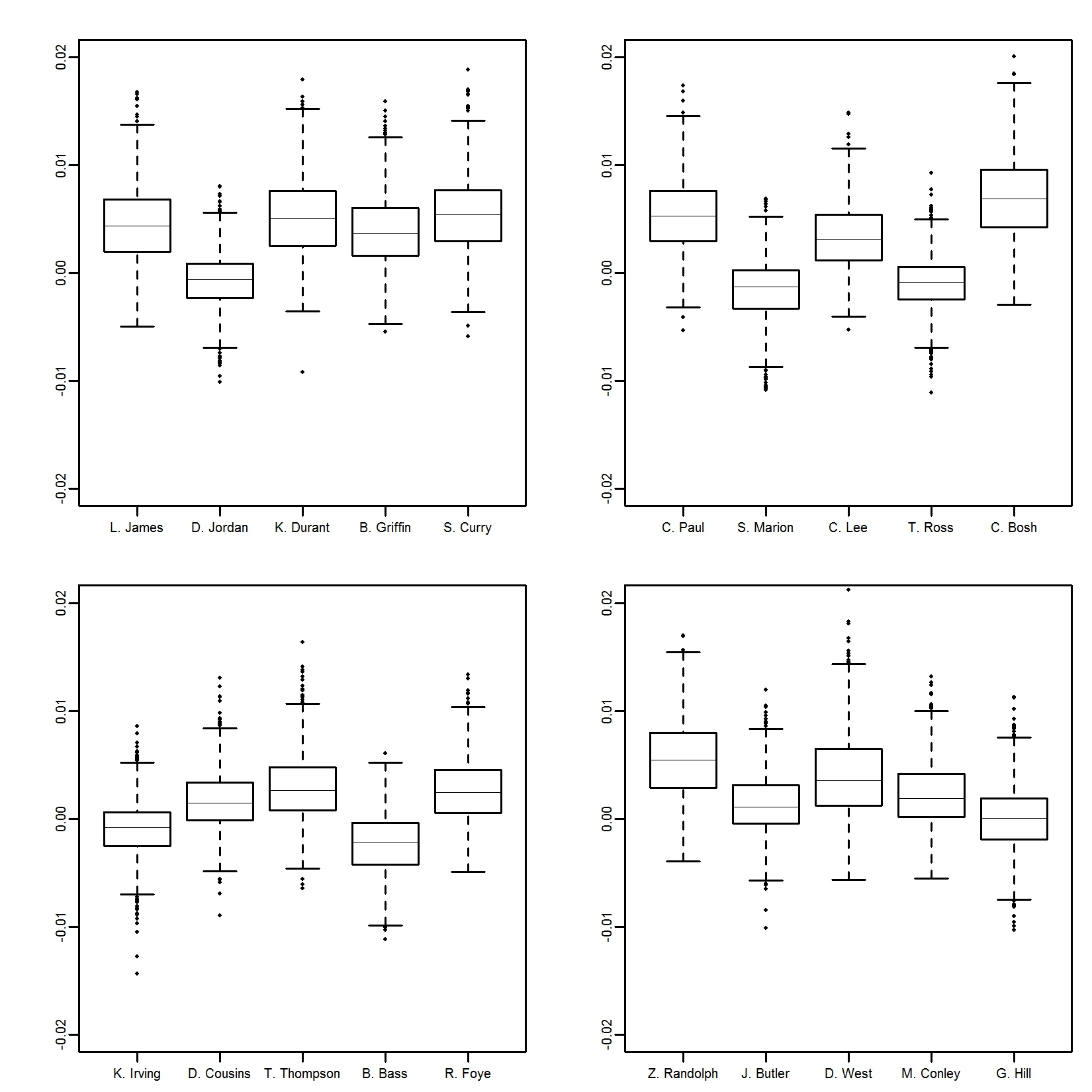}
\caption{Comparison box plots of partial effects of players with similar leverage profiles}
\label{fig:player_compare}
\end{figure}

Directly comparing partial effects for all pairs of players is complicated by the fact that players perform in different contexts.
To determine which players are most comparable, we determine the total number of shifts each player played, his team's average win probability at the start of these shifts, the average duration of these shifts, and the average length of each shift.
We call this information a player's \textit{leverage profile}.
We then compute the Mahalanobis distance between the leverage profiles of each pair of players.
Table~\ref{tab:player_compare} shows the four players with the most similar leverage profile for several players and Figure~\ref{fig:player_compare} shows comparison box plots of the posterior distribution of their partial effects.

\begin{table}[!h]
\centering
\caption{Most similar leverage profiles. Mahalanobis distance shown in parentheses}
\label{tab:player_compare}
\begin{tabular}{l|ll}
\hline
Player & ~ Similar Players & \\ \hline
LeBron James & DeAndre Jordan (0.025) & Kevin Durant (0.055)  \\
~ & Blake Griffin (0.082) & Stephen Curry (0.204)  \\ \hline
Chris Paul & Shawn Marion (0.081) & Courtney Lee (0.103) \\
~ & Terrence Ross (0.126) & Chris Bosh (0.141) \\ \hline
Kyrie Irving & DeMarcus Cousins (0.080) & Tristan Thompson (0.087) \\
~ & Brandon Bass (0.099) & Randy Foye (0.109) \\ \hline
Zach Randolph & Jimmy Butler (0.020) & David West (0.045) \\
~ & Mike Conley (0.063) & George Hill (0.073) \\\hline
\end{tabular}
\end{table}

We see that the posterior distributions of partial effects for each player in Table~\ref{tab:player_compare} are well-separated from the posterior distribution of partial effects of the player with the most similar leverage profile. 
For instance, LeBron James' leverage profile is most similar to DeAndre Jordan's, but we see that James' posterior distribution is located to the right of Jordan's and we find that in 884 of the 1000 posterior samples, James' partial effect is greater than Jordan's.
This suggests that while James and Jordan played in similar contexts, James' performance in these situations was more helpful to his team than Jordan's.

\subsection{Team Effects}
\label{sec:team_effects}

\begin{figure}[!h]
\centering
\includegraphics[width = 4.5in]{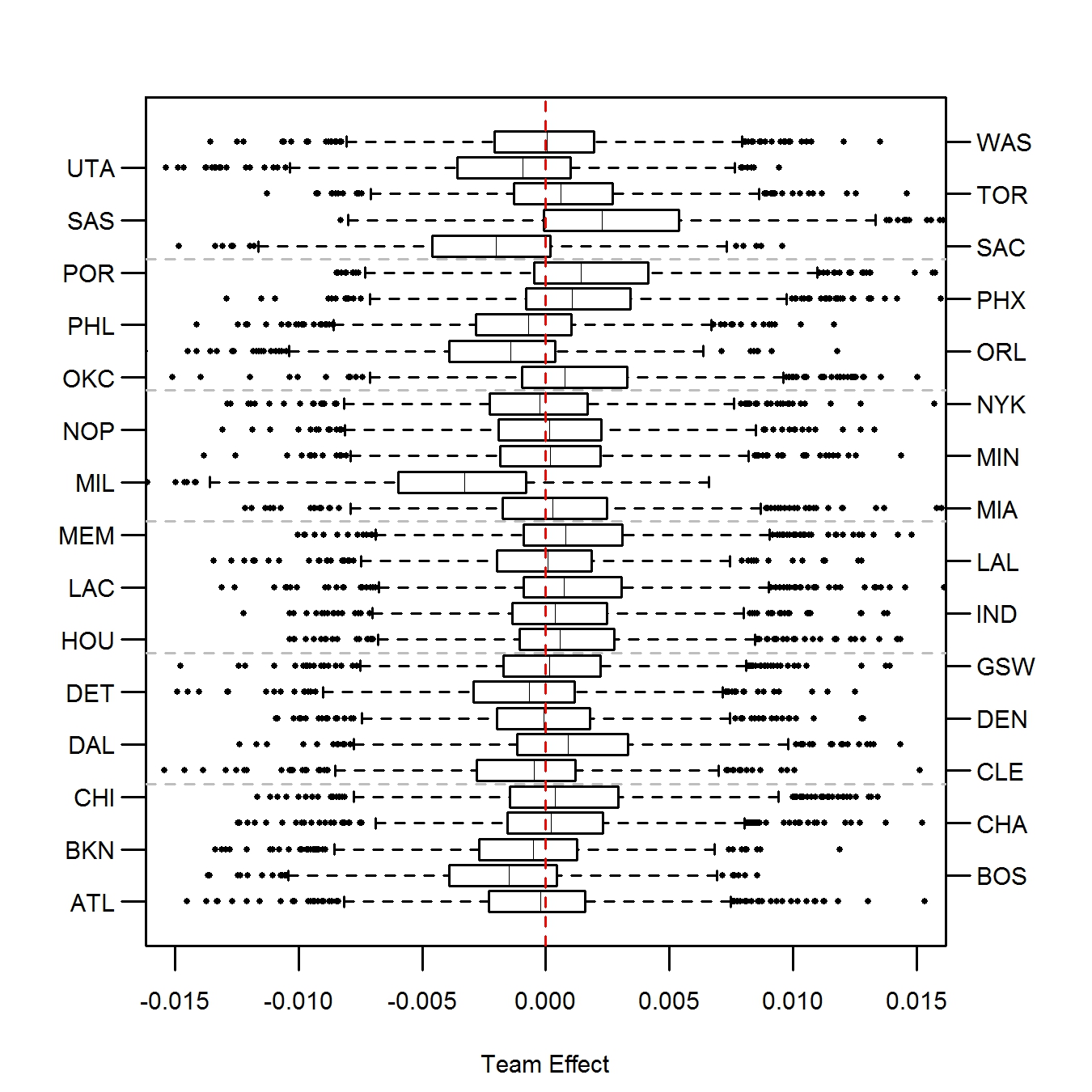}
\caption{Comparison box plots of the posterior distribution of team effects.}
\label{fig:team_effect}
\end{figure}

Recall that the inclusion of team effects, $\tau,$ in Equation~\ref{eq:player_team_model} was to ensure that the partial effects of players were not overly deflated if they played on bad teams or overly inflated if they played on good teams.
Figure~\ref{fig:team_effect} shows box plots of the posterior distribution of all team effects.

\begin{figure}[!h]
\centering
\includegraphics{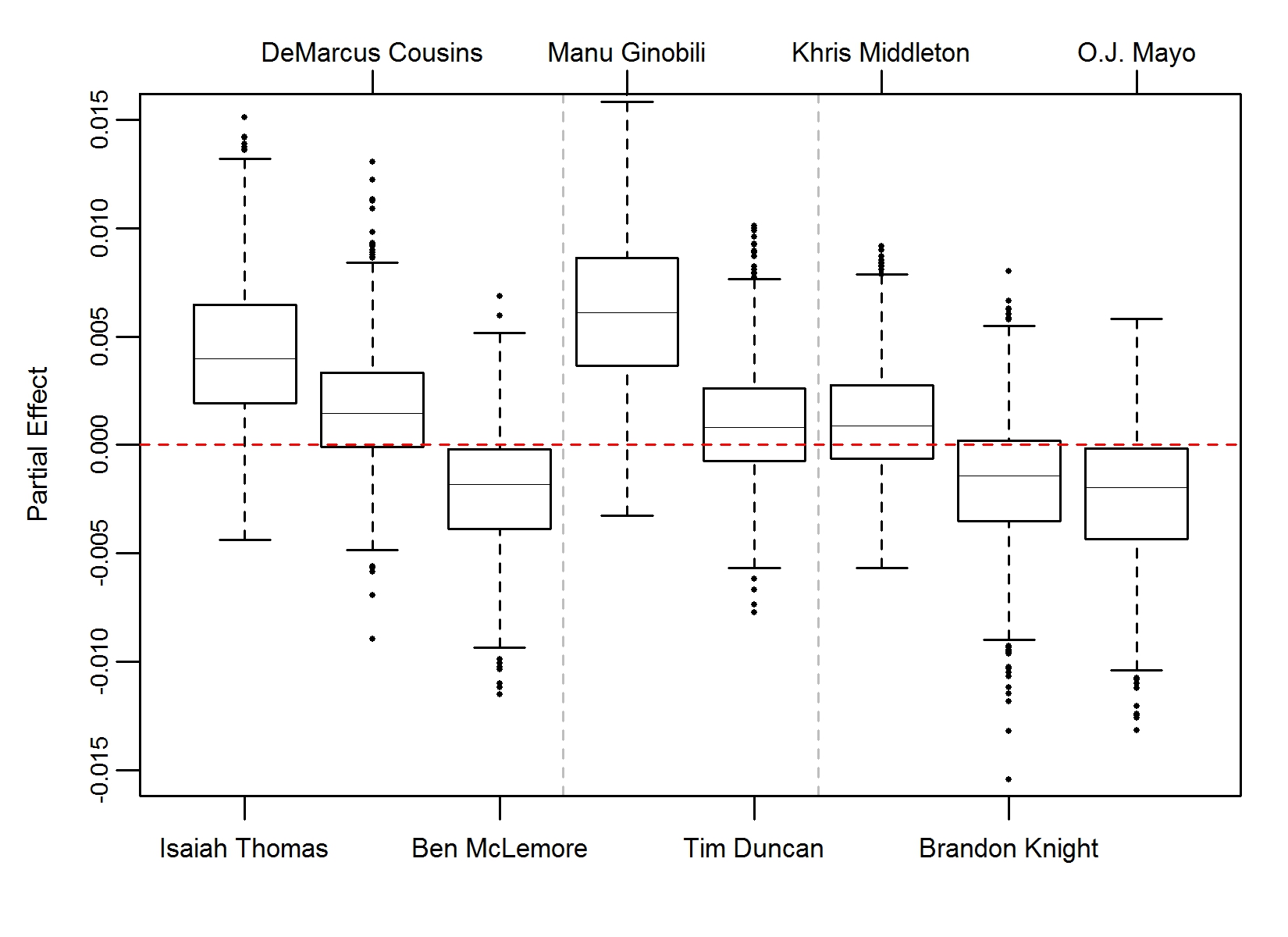}
\caption{Comparison box plots of partial effects of selected Bucks, Kings, and Spurs players}
\label{fig:bucks_kings_spurs_comparison}
\end{figure}

We see that the Milwaukee Bucks and Sacramento Kings have a noticeably negative effect.
This suggests that opposing teams generally increased their chances of winning, regardless of which five Bucks or Kings players were on the court.
This is in contrast with the San Antonio Spurs, whose team effect is substantially positive.
Figure~\ref{fig:bucks_kings_spurs_comparison} shows comparative box plots of the posterior distribution of the partial effects for a few Bucks, Kings, and Spurs players.

The fact that the posterior distributions of Isaiah Thomas', DeMarcus Cousins', and Khris Middleton's partial effects are predominantly concentrated on the positive axis indicates that their performance stood out despite the relatively poor quality of their team. 
On the other hand, the posterior distributions of Ben McLemore's, O.J. Mayo's, and Brandon Knight's partial effects are predominantly concentrated on the negative axis, indicating that their teams' already diminished chances of winning decreased when they were on the court. 
The fact that Manu Ginobili has such a large positive partial effect is especially noteworthy, given the Spurs' already large positive effect. 

\section{Impact Ranking}
\label{sec:impact_ranking}


Since we may view a player's partial effect as an indication of his value to his team, we can generate a rank-ordering of the players on each team based on their partial effects.
Intuitively, we could rank all of the members of a particular team by the posterior mean or median of their partial effects.
Such an approach, however, does not incorporate the joint uncertainty of the partial effects.
Alternatively, for each team and each posterior sample of $\theta,$ we could rank the partial effects of all players on that team and then identify the rank-ordering with highest posterior frequency.
Unfortunately, since there are over one trillion orderings of 15 players (the minimum number of players per team), such an approach would require an impractical number of posterior samples. 
Instead, we propose to average the player rankings over the 1000 posterior samples to get their \textbf{Impact Ranking}.
Table~\ref{tab:impact_ranking} shows the Impact Ranking for the players on the  San Antonio Spurs and the Miami Heat, with the most common starting lineup bolded and players who played very limited minutes starred.

\begin{table}[!h]
\centering
\caption{Impact Ranking for San Antonio Spurs and Miami Heat players. For each player, we report both his salary and the approximate probability that his partial effect is greater than the that of the player ranked immediately after him. Starred players played very limited minutes.}
\label{tab:impact_ranking}
\begin{tabular}{llcl}
\\
\hline
Rank&San Antonio Spurs&~~~~~~~~~&Miami Heat\\
\hline
1. & Manu Ginobili (\$7.5M, 0.719)& &\textbf{Chris Bosh} (\$19.1M, 0.650)\\
2. & \textbf{Danny Green} (\$3.8M, 0.540) & & \textbf{LeBron James} (\$19.1M, 0.683) \\
3. &Patty Mills (\$1.3M, 0.531)& & \textbf{Mario Chalmers} (\$4M, 0.577)\\
4. &\textbf{Kawhi Leonard} (\$1.9M, 0.631) & & Ray Allen (\$3.2M, 0.571)\\
5. &\textbf{Tiago Splitter} (\$10M, 0.510)& & Toney Douglas (\$1.6M, 0.486) \\
6. &\textbf{Tony Parker}  (\$12.5M, 0.554) & & Roger Mason Jr. (\$0.8M, 0.509) \\
7. & \textbf{Tim Duncan} (\$10.4M, 0.518)& & Chris Andersen (\$1.4M, 0.518)\\
8. & Damion James$^{*}$ (\$20K, 0.489)& & James Jones (\$1.5M, 0.520)\\
9. & Boris Diaw (\$4.7M, 0.566)& & \textbf{Dwyane Wade} (\$18.7M, 0.515) \\
10. & Matt Bonner (\$3.9M, 0.582)& & DeAndre Liggins$^{*}$ (\$52K, 0.520)\\
11. & Jeff Ayres (\$1.8M, 0.556)& & Norris Cole (\$1.1M, 0.570) \\
12. & Nando de Colo (\$1.4M, 0.561)& & Justin Hamilton$^{*}$ (\$98K, 0.541) \\
13. & Austin Daye (\$0.9M, 0.530)& & Michael Beasley (\$0.8M, 0.511)\\
14. & Aron Baynes (\$0.8M, 0.513)& & Greg Oden (\$0.8M, 0.503)\\
15. & Cory Joseph (\$1.1M, 0.583)& & Rashard Lewis (\$1.4M, 0.525)\\
16. & Marco Belinelli (\$2.8M)& & \textbf{Shane Battier} (\$3.3M, 0.618) \\
17. & & & Udonis Haslem (\$4.3M)\\
\hline
\end{tabular}
\end{table}

In Table~\ref{tab:impact_ranking}, we see that the most impactful player for the Spurs, Manu Ginobili, is a bench player, while five of the next six most impactful players were the most common starters.
This is in contrast to the Heat, for whom we only observe three starters in the top five most impactful players and a rather significant drop-off down to the remaining starters.
For instance, Dwayne Wade was not nearly as impactful as several Heat bench players and Shane Battier was even less valuable than several players who had very limited minutes (DeAndre Liggins and Justin Hamilton) or limited roles (Greg Oden). 
This indicates that the Heat did not rely much on Wade or Battier to win games, despite their appearance in the starting lineup. 
We can further compare each player's salary to his impact ranking to get a sense of which players are being over- or under-valued by their teams.
For instance, Patty Mills earned only \$1.3M, the eleventh highest salary on the Spurs, despite being the third most impactful player on the team. 
In contrast, Wade was the ninth most impactful player on Heat, despite earning nearly \$15 million dollars more than Mario Chalmers, who was the third most impactful player for the Heat.


\section{Impact Score}
\label{sec:impact_score}

A natural use of any player evaluation methodology is to generate a single ranking of all players and we could simply rank all players in the league according to the posterior mean of their partial effects.
Unfortunately, since the mean by influenced by a few very extreme value, such a ranking can overvalue players whose partial effects have large posterior variance.
To try to account for the joint variability of player effects, we can rank the players' partial effect estimates in each of our 1000 simulated posterior samples.
Then we could compute 95\% credible intervals for each player's partial effects-based rank.
We find, however, that these intervals are rather long. 
For instance, we find that LeBron James had the largest partial effect among all players in only 11 of the 1000 posterior samples and the 95\% credible interval for his rank is $[3, 317].$
Similarly, we find that Kevin Durant also had the largest partial effect among all players in 11 of the 1000 posterior samples and the 95\% credible for his rank is $[2, 300].$
It turns out that Dirk Nowitzki had the largest partial effect in the most number of posterior samples (39 out of 1000) but the credible interval for his rank is $[1, 158].$
Given the considerable overlap in the posterior distributions of player partial effects as seen in Figure~\ref{fig:player_effect_density}, it is not surprising to see the large joint posterior variability in player partial effects reflected in the rather long credible intervals of each player's partial effects-based ranks. 

We instead propose to rank players according to their \textbf{Impact Score}, which we define as the ratio between the posterior mean and the posterior standard deviation of a player's partial effect.
This definition is very similar to the Sharpe Ratio used in finance to examine the performance of an investment strategy.
We may view Impact Score as a balance between a player's estimated ``risk'' (i.e. uncertainty about his partial effect) and a player's estimated ``reward'' (i.e. average partial effect).
As an example, we find that the posterior mean of Iman Shumpert's partial effect is less than the posterior mean of Chris Bosh's partial effect (0.0063 compared to 0.0069).
We also find that the posterior standard deviation of Shumpert's partial effect is 0.0034 while it is 0.0039 for Bosh.
Between the two players, Shumpert gets the edge in Impact Score rankings because we are less uncertain about his effect, despite him having a smaller average effect compared to Bosh.
Table~\ref{tab:impact_score} shows the thirty players with largest Impact Scores.
Somewhat unsurprisingly, we see a number of superstars in Table~\ref{tab:impact_score}.
Patrick Patterson is a notable standout; as \cite{Cavan2014} and \cite{Lapin2014} note, he provided valuable three-point shooting and rebounding off the bench for the Toronto Raptors.

\begin{table}[!h]
\centering
\caption{Players with the highest Impact Scores}
\label{tab:impact_score}
\begin{tabular}{lll}
\\
\hline
1. Dirk Nowitzki (2.329)& ~~~~~~ & 16. Eric Bledsoe (1.274) \\
2. Patrick Patterson (1.939)& ~~~~~~ & 17. Dwight Howard (1.273)   \\
3. Iman Shumpert (1.823) & ~~~~~~ & 18. Danny Green (1.214)\\
4. Chris Bosh (1.802) & ~~~~~~ & 19. Deron Williams (1.212)  \\
5. Manu Ginobili (1.779) & ~~~~~~ & 20. Matt Barnes (1.206)  \\
6. James Harden (1.637) & ~~~~~~ & 21. Roy Hibbert (1.205) \\
7. Chris Paul (1.588) & ~~~~~~ & 22. J.J. Redick (1.201)  \\
8. Zach Randolph (1.56) & ~~~~~~ & 23. Shaun Livingston (1.201) \\
9. Joakim Noah (1.555) & ~~~~~~ & 24. Marcin Gortat (1.185)\\
10. Stephen Curry (1.514) & ~~~~~~ & 25. Greivis Vasquez (1.175)  \\
11. Nene Hilario (1.474) & ~~~~~~ & 26. Blake Griffin (1.174)\\
12. Andre Iguodala (1.445) & ~~~~~~ & 27. Anthony Tolliver (1.151) \\
13. Kevin Durant (1.410) & ~~~~~~ & 28. LaMarcus Aldridge (1.140) \\
14. LeBron James (1.324) & ~~~~~~ & 29. Courtney Lee (1.131) \\
15. Isaiah Thomas (1.310) & ~~~~~~ & 30. Nate Robinson (1.126) \\
\hline
\end{tabular}
\end{table}

It is important to note that our reported Impact Scores are subject to some degree of uncertainty, since we have to estimate the posterior mean and standard deviation of each player's partial effect.
This uncertainty amounts to MCMC simulation variability and induces some uncertainty in the reported player rankings. 
In order to quantify the induced uncertainty explicitly, we could run our sampler several times, each time generating a draw of 1000 simulated posterior samples and ranking the players according to the resulting Impact Scores.
We could then study the distribution of each player's ranking. 
While straightforward in principle, the computational burden of running our sampler sufficiently many times is rather impractical.
Moreover, we suspect the simulation-to-simulation variability in Impact Scores is small.
Since we are estimating the posterior mean and standard deviation of player partial effects with 1000 samples, we are reasonably certain that the estimated values are close to the true values. 
As a result, our reported Impact Scores are reasonably precise and we do not expect much variation in the player rankings.

\subsection{Comparison of Impact Score to Other Metrics}
\label{sec:comparison}

\begin{figure}[!h]
\centering
\includegraphics{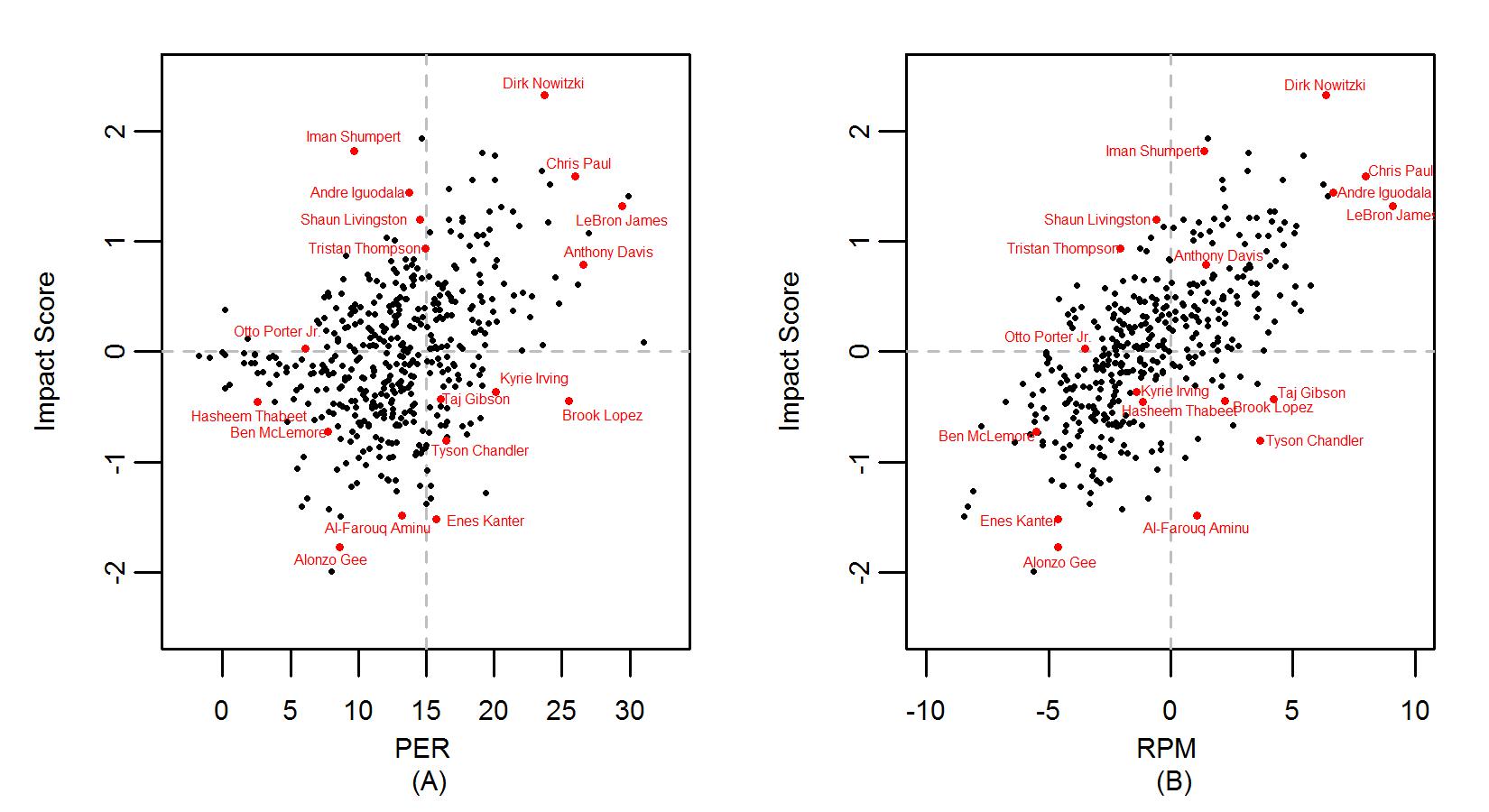}
\caption{Comparison of Impact Score to PER (A) and RPM (B). 
We find that RPM is much more consistent with Impact Score than is PER, though there are still several inconsistencies in overall player evaluation. Note that PER is calibrated so the league average is 15.00}
\label{fig:impact_comparison}
\end{figure}

\cite{Hollinger2004} introduced Player Efficient Rating (PER) to ``sum up all [of] a player's positive accomplishments, subtract the negative accomplishments, and a return a per-minute rating of a player's performance.''
Recently, ESPN introduced Real Plus-Minus (RPM) which improves on Adjusted Plus-Minus through a proprietary method that, according to \cite{Ilardi2014}, uses ``Bayesian priors, aging curves, score of the game and extensive out-of-sample testing.''
Figure~\ref{fig:impact_comparison} shows Impact Score plotted against PER and RPM.  
We note that of the 488 players in our data set, RPM was available for only 437.
In Figure~\ref{fig:impact_comparison}, we have excluded the six players whose PER is greater than 33 or less than -3 so that the scale of the figure is not distorted by these extreme values.

We find that the correlation between Impact Score and PER is somewhat moderate (correlation 0.226) and that Impact Score is much more highly correlated with RPM (correlation of 0.655). 
This is somewhat expected, since PER is essentially context-agnostic and RPM at least partially accounts for the context of player performance.
To see this, we note that the number of points a player scores is a key ingredient in the PER computation. 
What is missing, however, is any consideration of \textit{when} those points were scored.
RPM is more context-aware, as it considers the score of the game when evaluation player performance.
However, since the RPM methodology is proprietary, the extent to which the context in which a player performs influences his final RPM value remains unclear.

\begin{figure}[!h]
\centering
\includegraphics{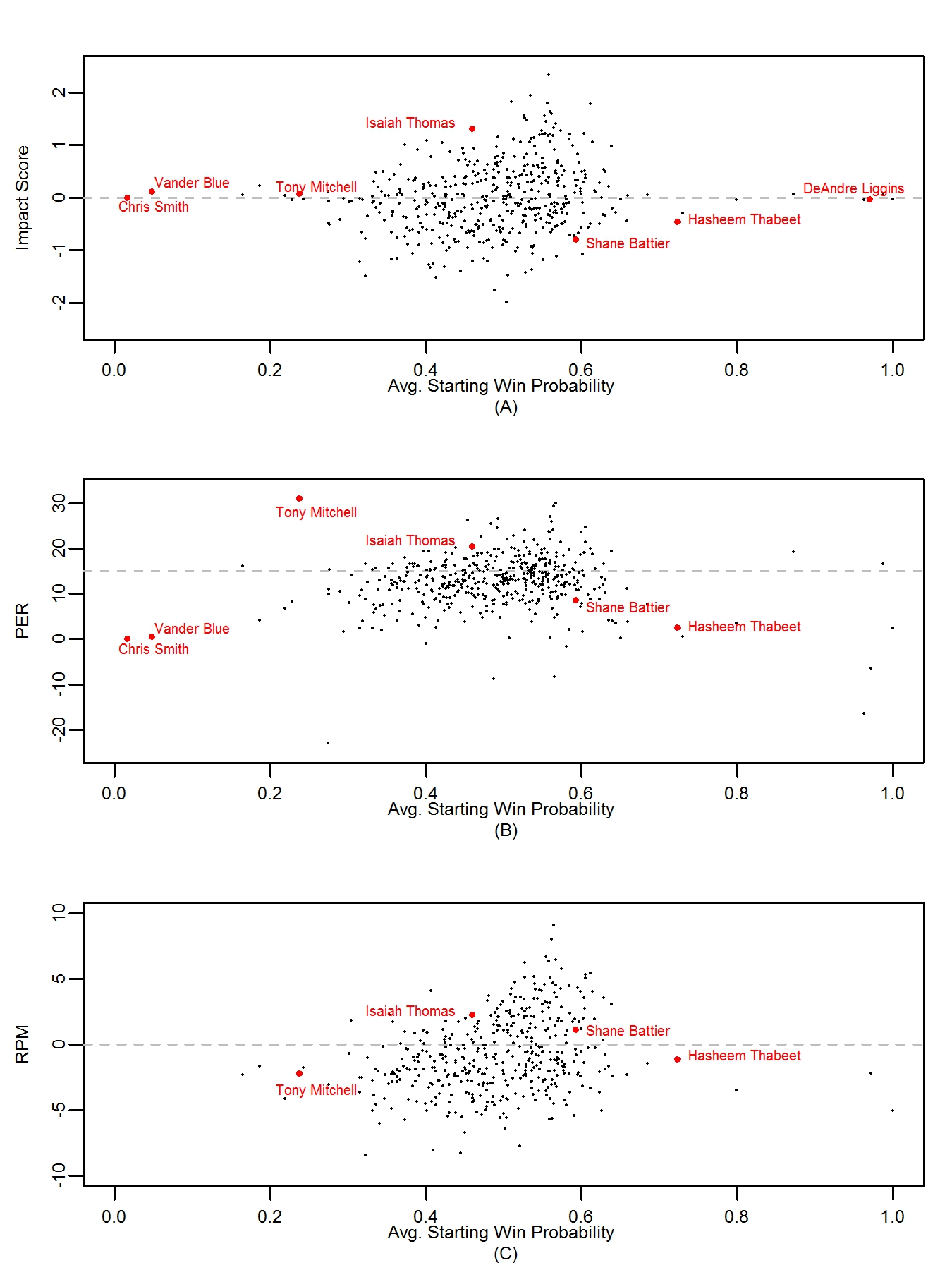}
\caption{Impact Score, PER, and RPM plotted against average starting win probability. Note that RPM was unavailable for 51 players.}
\label{fig:start_winProb_comparison}

\end{figure}

As we noted in Section~\ref{sec:introduction}, metrics like PER and point-differential metrics can overvalue low-leverage performances.
An extreme example of this is DeAndre Liggins' PER of 129.47.
Liggins played in a single game during the 2013-14 regular season and in his 84 seconds of play, he made his single shot attempted and recorded a rebound.
We note, however, that Liggins entered the game when his team had a 96.7\% chance of winning the game and his performance did not improve his team's chances of winning in any meaningful way. 
Figure~\ref{fig:start_winProb_comparison} plots each player's Impact Score, PER, and RPM against the average win probability of each player's shifts.
In Figure~\ref{fig:start_winProb_comparison}, we have included the players with very negative PER values who were excluded from Figure~\ref{fig:impact_comparison}.

In Figure~\ref{fig:start_winProb_comparison}, we see that the average starting win probability for Chris Smith, Vander Blue, Tony Mitchell, and DeAndre Liggins was less than 0.2 or greater than 0.8, suggesting that they played primarily in low-leverage situations.
We see that while their PERs ranged from -23 to 130, their Impact Scores are all very close to zero.
This confirms that our methodology correctly values so-called ``garbage time'' performance.
It is interesting to note Hasheem Thabeet played when his team had, on average, above a 70\% of winning the game.
His negative Impact Score is an indication that his performance generally hurt his team's chances of winning and we find that he had a negative partial effect in 678 of the 1000 posterior samples.

While it is encouraging that there is at least some positive correlation between Impact Scores and PER, simply looking at the correlation is not particularly informative, as these metrics are measuring rather different quantities. 
Of greater interest, perhaps, is to see when PER and Impact Score agree and when they disagree. 
For instance, we find players like LeBron James, Chris Paul and Dirk Nowitzki who have both large PER values and large Impact Scores.
The large PER values are driven by the fact that they efficiently accumulated more positive box-score statistics (e.g. points, assists, rebounds, etc.) than negative statistics (e.g. turnovers and fouls) and the large Impact Scores indicate that their individual performances helped improve their team's chances of winning.
On the other hand, Brook Lopez and Kyrie Irving have the ninth and twenty-ninth largest PER values but their rather middling Impact Scores suggest that, despite accumulating impressive individual statistics, their performances did not actually improve their teams' chances of winning.

In contrast to Irving and Lopez, players like Iman Shumpert and Andre Iguodala have below-average PER values but rather large Impact Scores.
Shumpert has a PER of 9.66, placing him in the bottom 25\% of the league, but has the fourth largest Impact Score.
This suggests that even though Shumpert himself did not accumulate particularly impressive individual statistics, his team nevertheless improved its chances of winning when he was on the court.
It is worth noting that Shumpert and Iguodala are regarded as top defensive players. 
As \cite{GoldsberryWeiss2013} remark, conventional basketball statistics tend to emphasize offensive performance since there are not nearly as many discrete defensive factors to record in a box score as there are offensive factors.
As such, metrics like PER can be biased against defensive specialists.
It is re-assuring, then, to see that Impact Score does not appear to be as biased against defensive players as PER.

It is important to note that the fact that Shumpert and Iguodala have much larger Impact Scores than Lopez and Irving does not mean that Shumpert and Iguodala are inherently better players than Lopez and Irving.
Rather, it means that Shumpert's and Iguodala's performances helped their teams much more than Irving's or Lopez's. 
One explanation for the discrepancies between Lopez and Irving's Impact Scores and PERs could be coaching decisions.
The fact that Lopez and Irving were accumulating impressive individual statistics without improving their respective teams' chances of winning suggests that their coaches may not have been playing them at opportune times for their teams. 
In this way, when taken together with a metric like PER, Impact Score can provide a more complete accounting and evaluation of a player's performance.

\subsection{Year-to-year correlation of Impact Score}
\label{sec:year_to_year_impactScore}

\begin{figure}[!h]
\centering
\includegraphics[width = 4.5in]{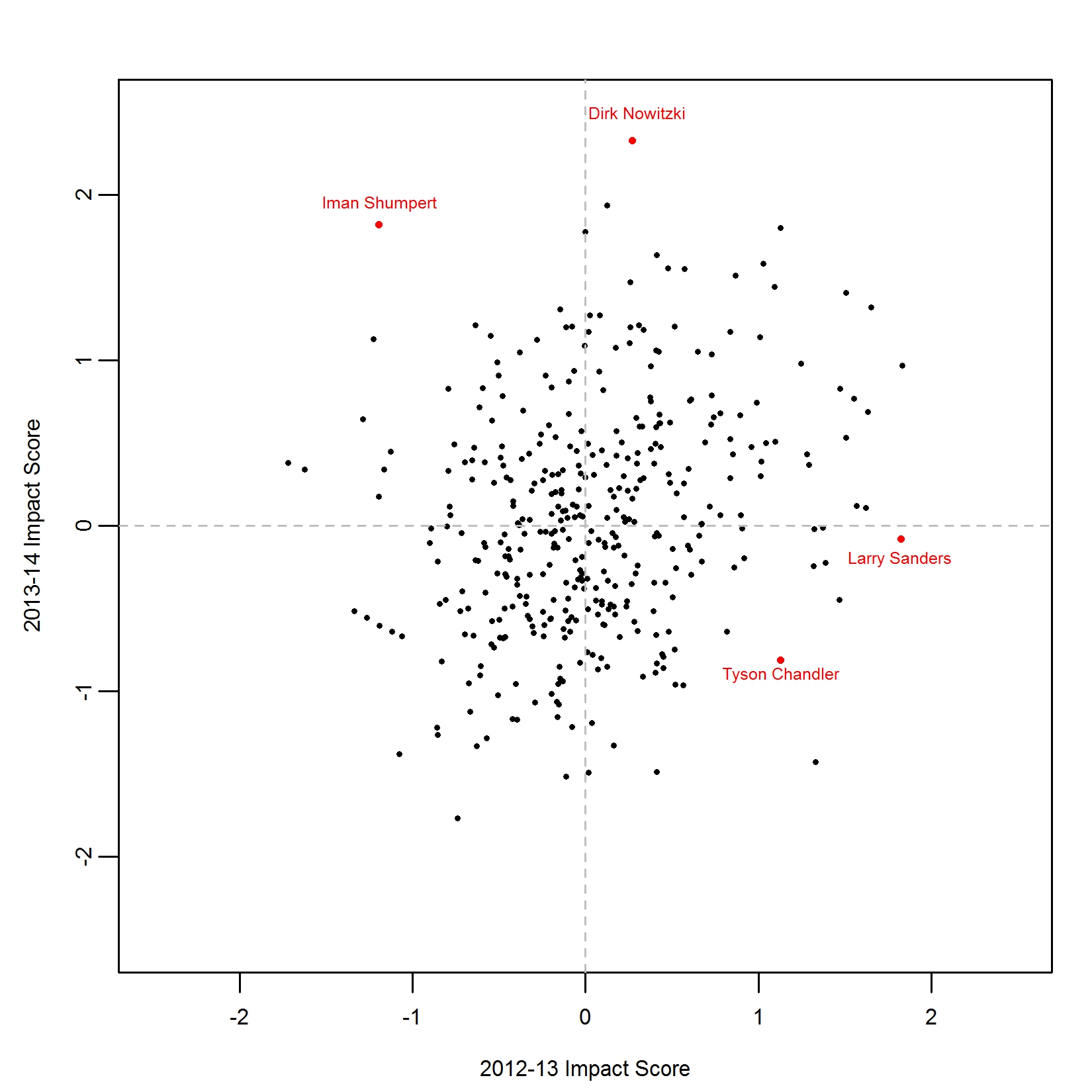}
\caption{Impact Scores in 2012 and 2013}
\label{fig:impact2012_2013} 
\end{figure}

A natural question to ask about any player evaluation metric is how stable it is year-to-year. 
In other words, to what extent can we predict how a player ranks with respect to one metric in a season given his ranking in a previous season.
Using play-by-play data from the 2012-13 regular season, we can fit a model similar to that in Equation~\ref{eq:player_team_model} and compute each player's Impact Score for that season. 
There were 389 players who played in both the 2012-13 and 2013-14 seasons and Figure~\ref{fig:impact2012_2013} plots there players' 2012-13 Impact Scores against their 2013-14 Impact Scores.

We observe that the correlation between 2012-13 and 2013-2014 Impact Score is 0.242, indicating a rather moderate positive trend. 
We notice, however, that there are several players whose Impact Scores in 2012-13 are much different than their Impact Scores in 2013-14.
For instance, Iman Shumpert's and Dirk Nowitzki's Impact Scores increased dramatically between the two season.
At the other end of the spectrum, players like Larry Sanders and Tyson Chandler displayed sharp declines in their Impact Scores.
On further inspection, we find that all of these players missed many games due to injury in the seasons when they had lower Impact Scores.
Upon their return from injury, they played fewer minutes while they continued to rehabilitate and re-adjust to playing at a high-level.
In short, the variation in the \textit{contexts} in which these players performed is reflected in the the season-to-season variation in their Impact Score.

\begin{figure}[!h]
\centering
\includegraphics[width = 4.5in]{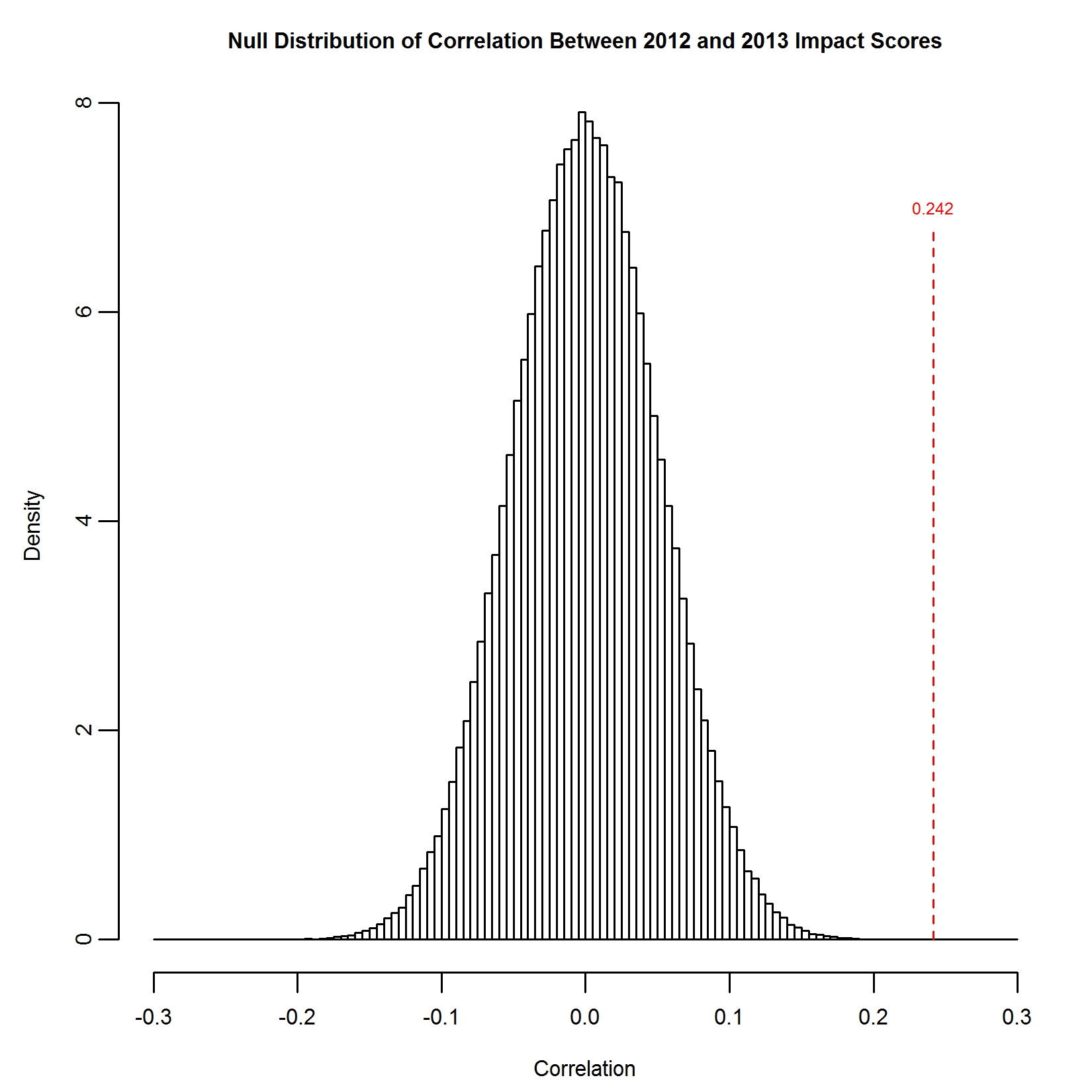}
\caption{Null distribution of correlation between 2012-13 and 2013-14 Impact Scores under the hypothesis that they are independent. The observed correlation of 0.242 is shown in red}
\label{fig:impactScore_cor2012_2013}
\end{figure}

Because it is context-dependent, we would not expect the year-to-year correlation for Impact Scores to be nearly as high as the year-to-year correlation for PER (correlation of 0.75), which attempts to provide a context-agnostic assessment of player contribution.
Nevertheless, we may still assess the significance of the correlation we have observed using a permutation test.
To simulate the distribution of the correlation between 2012-13 and 2013-14 Impact Scores, under the hypothesis that they are independent, we repeatedly permute the observed 2013-14 Impact Scores and compute the correlation between these permuted scores and the observed 2012-13 Impact Scores. 
Figure~\ref{fig:impactScore_cor2012_2013} shows a histogram of this null distribution based on 500,000 samples.

We find that the observed correlation is significantly different than zero. 
This indicates that even though Impact Scores are inherently context-dependent, a player's Impact Score is one season is moderately predictive of his Impact Score in the next, barring any significant changes in the contexts in which he plays.

\subsection{Multi-season Impact Score}

Though the context in which players perform between seasons can be highly variable, it is arguably more stable across multiple seasons. 
In light of this, we can re-fit our models using all of the play-by-play data from 2008-09 to 2010-11 and from 2011-12 to 2013-14, and estimated each player's partial effect separately in both time period. 
Note that for each season considered, the change in win probability during a shift was estimated using data from all prior seasons.

Somewhat surprisingly, we find that the posterior standard deviations of the player partial effects estimated over multiple seasons is not substantially smaller than when we consider one seasons at a time, despite having much more data.
For instance, the posterior standard deviation of LeBron James' partial effect in the 2013-14 season is 0.0035 while it is 0.002 over the three season span from 2008-09 to 2010-11.
Table~\ref{tab:multi_year_impact} shows the top 10 Impact Scores over these two three-season periods.

\begin{table}[!h]
\centering
\caption{Impact Score computed over three seasons windows}
\label{tab:multi_year_impact}
\begin{tabular}{ll} \hline
2008-09 -- 2010-11 & 2011-12 -- 2013-14 \\ \hline
LeBron James (5.400) & LeBron James (3.085)  \\
Dirk Nowitzki (3.758) & Chris Paul (3.041)\\ 
Chris Paul (3.247) & Amir Johnson (2.982)\\
Dwyane Wade (2.948) & Stephen Curry (2.919) \\
LaMarcus Aldridge (2.775) & Andre Iguodala (2.805)\\
Steve Nash (2.770) & Mike Dunleavy (2.790)\\
Tim Duncan (2.679) & Dirk Nowitzki (2.733) \\
Matt Bonner (2.178) & Kevin Durant (2.426) \\
Kevin Garnett (2.125) & Paul George (2.332)\\
\hline
\end{tabular}
\end{table}

Quite clearly, LeBron James stands out rather prominently, especially in the 2008-2010 time period, as far and away the most impactful player over those three seasons.
We note that James' 2013-2014 Impact Score is much less than either of his multi-season Impact Scores.
This indicates that while James may have been most impactful player over the course of several seasons, in that particular season, he was not as impactful.

\begin{figure}[!h]
\centering
\includegraphics[width = 4.5in]{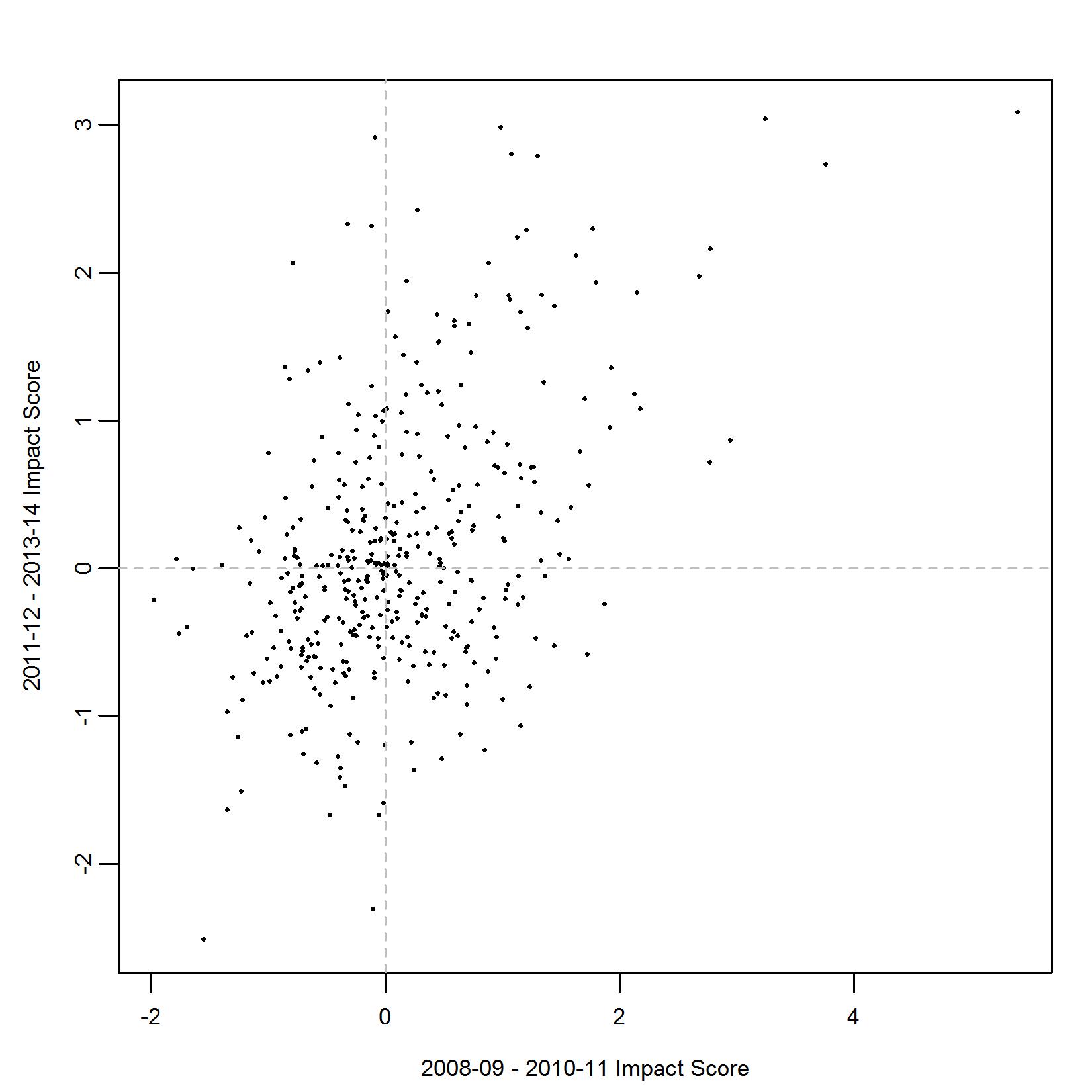}
\caption{Impact Scores computed over 2008-2010 and 2011-2013}
\label{fig:multi_year_impact}
\end{figure}

Figure~\ref{fig:multi_year_impact} shows the Impact Scores from 2011-2013 plotted against the Impact Scores 2008-2010.
The correlation between these scores is 0.45, which is larger than the year-to-year correlation in Impact Score.
The players with discordant single season Impact Scores highlighted in Figure~\ref{fig:impact2012_2013} were all recovering from significant injuries that required them to miss many games and play restricted minutes for a good portion of the season. 
Since there are generally few injuries which span significant portions of multiple seasons, the context in which players perform tend to stabilize across several seasons. 

\section{Lineup comparison}
\label{sec:lineup_comparison}

As a further study of the full covariance structure of $\theta$ and $\tau,$ we can compare how different five-man lineups match up against each other. 
To simulate the posterior distribution of a five-man lineup's effect on its team's win probability, we simply sum the corresponding entries of each posterior sample of $\theta.$
With these samples, we can compute each lineup's Impact Score just as we did for player's in Section~\ref{sec:impact_score}: we divide the posterior mean of the lineup's effect by the posterior standard deviation of its effect. 
Table~\ref{tab:lineup_impact_score} shows the ten lineups with the largest Impact Scores.

\begin{table}[!h]
\centering
\caption{Lineups with the largest impact score. }
\label{tab:lineup_impact_score}
\begin{tabular}{lll}
\\
\hline
~~~~Lineup & Impact Score & Minutes\\
\hline
1.~ Stephen Curry, Klay Thompson, Andre Iguodala & 2.98 & 780.25 \\
~~~~ David Lee, Andrew Bogut & ~ & ~ \\ \hline 
2.~Chris Paul, J.J. Redick, Matt Barnes & 2.88 & 88.57 \\
~~~~Blake Griffin, DeAndre Jordan & & \\ \hline
3.~Stephen Curry, Klay Thompson, Andre Iguodala & 2.82 & 31.75 \\
~~~~David Lee, Jermaine O'Neal & & \\ \hline
4.~George Hill, Lance Stephenson, Paul George & 2.58 & 1369.38 \\
~~~~David West, Roy Hibbert & & \\ \hline 
5.~Mario Chalmers, Ray Allen, LeBron James & 2.57 & 34.28 \\
~~~~Chris Bosh, Chris Andersen & & \\ \hline
6.~Patrick Beverley, James Harden, Chandler Parsons & 2.51 & 589.97 \\
~~~~Terrence Jones, Dwight Howard & & \\ \hline 
7.~Mario Chalmers, Dwyane Wade, LeBron James & 2.46 & 26.2 \\
~~~~Chris Bosh, Chris Andersen & & \\ \hline
8.~C.J. Watson, Lance Stephenson, Paul George & 2.42 & 118.27 \\
~~~~David West, Roy Hibbert & & \\ \hline
9.~John Wall, Bradley Beal, Trevor Ariza & 2.38 & 384.03 \\
~~~~Nene Hilario, Marcin Gortat & & \\ \hline
10.~Patrick Beverley, James Harden, Chandler Parsons & 2.38 & 65.58 \\
~~~~Donatas Motiejunas, Dwight Howard & & \\ \hline
\end{tabular}
\end{table}

\begin{figure}[!h]
\centering
\includegraphics[width = 4.5in]{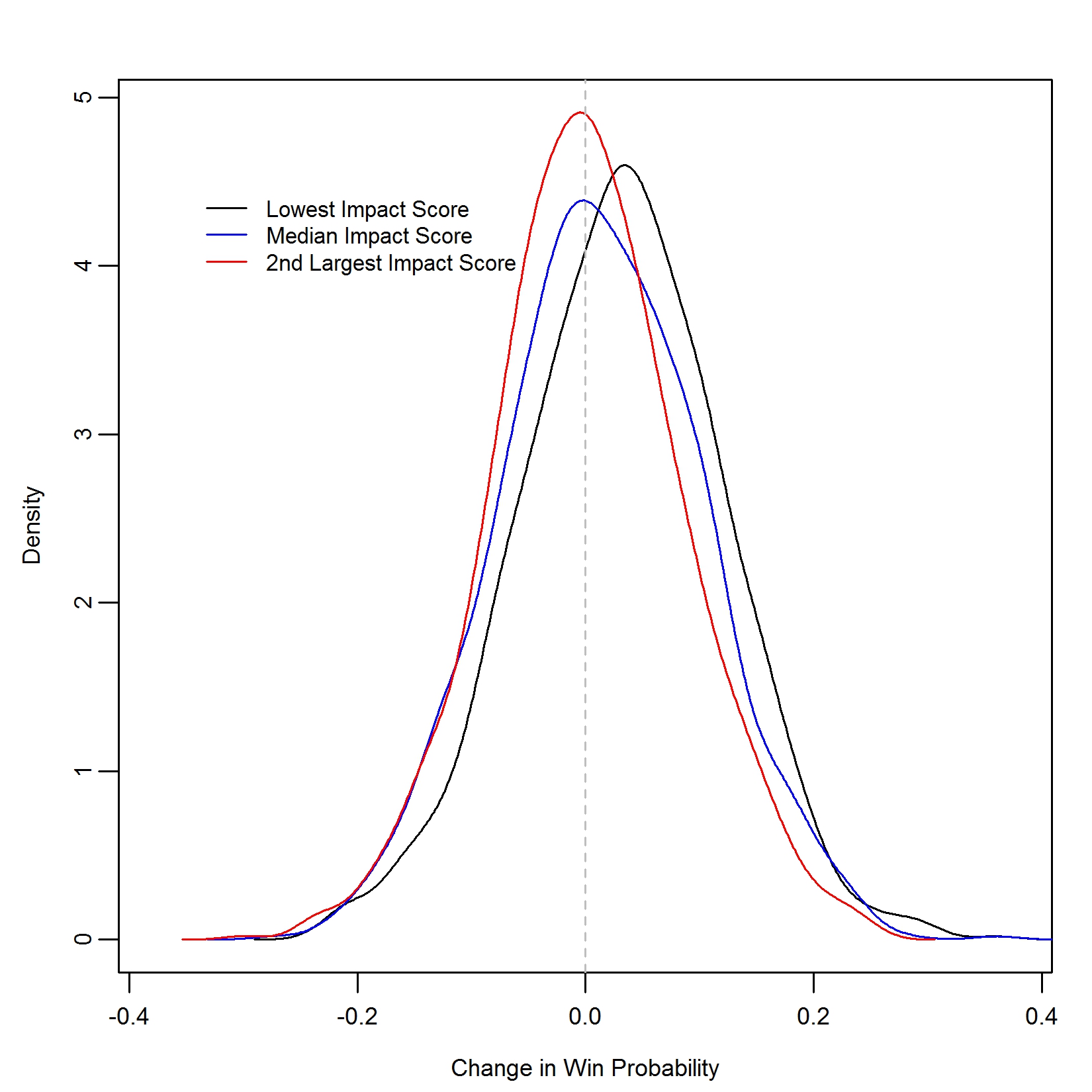}
\caption{Posterior predictive density in win probability of the lineup with the largest impact score matched up with three other lineups}
\label{fig:lineup_compare}
\end{figure}

We can also simulate draws from the posterior predictive distribution of the change in home team win probability for each home/away configurations of two five-man lineups using our posterior samples of $\left(\mu, \theta, \tau, \sigma^{2}\right).$
For a specific home/away configuration, we construct vectors of signed indicators, $\mathbf{P}^{*}$ and $\mathbf{T}^{*}$, to encode which players and teams we are pitting against one another.
For each sample of $\left(\mu, \theta, \tau, \sigma^{2}\right)$ we compute
$$
\mu + \mathbf{P}^{*\top}\theta + \mathbf{T}^{*\top}\tau + \sigma z
$$
where $z \sim N(0,1),$ to simulate a sample from the posterior predictive distribution of the change in the home team's win probability for the given matchup.
In particular, we consider pitting the lineup with the largest Impact Score (Stephen Curry, Klay Thompson, Andre Iguodala, David Lee, Andrew Bogut) against three different lineups:  the lineup with second largest Impact Score (Chris Paul, J.J. Redick, Matt Barnes, Blake Griffin, DeAndre Jordan), the lineup with the smallest Impact Score (Eric Maynor, Garrett Temple, Chris Singleton, Trevor Booker, Kevin Seraphin), and the lineup with the median Impact Score (Donald Sloan, Orlando Johnson, Solomon Hill, Lavoy Allen, Roy Hibbert).
The median lineup's Impact Score is the median of all lineup Impact Scores. 
Figure~\ref{fig:lineup_compare} shows the posterior predictive densities of the change in win probability in a single shift when the lineup with largest Impact Score plays at home.

Unsurprisingly, when the lineup with largest Impact Score is pitted against the lineup with smallest Impact Score, the predicted change in win probability is positive about 65\% of the time and is greater than 0.1 just over 23\% of the time.
It is also reassuring to see that the density corresponding to the matchup against the median lineup lies between the two extremes considered.
Rather surprisingly, however, we find that when the lineup with largest Impact Score is pitted against the lineup with second largest Impact Score, the change in win probability is negative about 55\% of the time.
We find that posterior mean effect of the Paul-Reddick-Barnes-Griffin-Jordan lineup is 0.0166 while the posterior mean effect of the Curry-Thompson-Iguodala-Lee-Bogut lineup is 0.0150.
The difference in Impact Score is driven by the difference in the posterior standard deviation of each lineup's effect (0.0050 for Curry-Thompson-Iguodala-Lee-Bogut and 0.0058 for Paul-Reddick-Barnes-Griffin-Jordan).
Because of the disparity in playing time (780.25 minutes vs 88.57 minutes), we are less uncertain about the effect of the Curry-Thompson-Iguodala-Lee-Bogut lineup and the additional certainty makes up for the smaller average effect.
This highlights an important feature of Impact Score: it tries to balance the estimated effect against the uncertainty in this estimate.

At this point, it is worth nothing that while the change in win probability over the course of any shift is constrained to lie between -1 and 1, none of our modeling assumptions restrict the range of the predicted change in win probability in any of the match-ups considered to lie in this range.
In particular, since we have a conditional normal model, it could be the case that $\sigma z$ term pushes our prediction outside of the interval $[-1,1].$
In light of this, it is reassuring to find that the support of posterior predictive distributions of the change in win probability in all of the match-ups considered is in $[-0.4, 0.4].$

\section{Discussion}
\label{sec:discussion}

In this paper, we have estimated each NBA player's effect on his team's chances of winning, after accounting for the contributions of his teammate and opponents.
By focusing on win probability, our model simultaneously down-weights the importance of performance in low-leverage (``garbage time'') and up-weights the importance of high-leverage performance, in marked contrast to existing measures like PER which provide context-agnostic assessments of player performance. 
Since our estimates of player effects depend fundamentally on the context in which players perform, our estimates and derived metrics are necessarily retrospective in nature.
As a result, our results do not display nearly as high of a year-to-year correlation as other metrics.
We would argue, however, that the somewhat lower year-to-year repeatability of our derived metrics are offset by the fact that they provide a much more complete accounting of how a player helped their teams win in a particular season.
When taken together with a metric like PER, our results enable us to determine whether the performance of a player who recorded impressive box-score totals actually improved his team's chances of winning the game.
Ultimately, our model and derived metrics serve as a complement to existing measures of player performance and enables us to contextualize individual performances in a way that existing metrics alone cannot. 

We have introduced a new method for estimating the probability that a team wins the game as a function of its lead and how much time is remaining. 
Our win probability estimates can be viewed as a middle-ground between the empirical estimates, which display extreme discontinuity due to small sample size issues, and existing probit regression model estimates, which do not seem to fit the empirical observations well. 
Though our win probability estimates are generally quite precise, our choice of smoothing window $[T - 3, T + 3] \times [L - 2, L + 2]$ is admittedly rather simplistic. 
This is most pronounced near the end of the game, when a single possession can swing the outcome and it less reasonable to expect the win probability when leading by 2 points is similar to the win probability when trailing by 2 points.
To deal with this, one could let the window over which we aggregate games vary with both time and lead instead of using a fixed window.
We also note that the choice of a hard threshold of $L = \pm 20$ in determining the number of pseudo-wins, $\alpha_{T,L},$  and pseudo-losses, $\beta_{T,L}$, to add is arbitrary and we could just as easily have selected $L = \pm 25$ or  $\pm 30.$
Alternatively, $\alpha_{T,L}$ and $\beta_{T,L}$ could be selected at random from a specified distribution depending on the time and lead or we can place a further hyper-prior on $\left(\alpha_{T,L}, \beta_{T,L}\right).$
Unfortunately, estimates from the first approach may not be reproducible and explicitly computing the Bayes estimator of $p_{T,L}$ in the second approach can be difficult. 
While a more carefully constructed prior can, in principle, lead to estimates that more accurately reflect our subjective beliefs about how win probability evolves, one must take care not to select a prior that can overwhelm the observed data.

Looking at our win probability estimates, we find that a unit change in time corresponds to a smaller change in win probability than a unit change in lead, especially near the end of close games.
This can introduce a slight bias against players who are frequently substituted into games on defensive possessions and taken out of the game on offensive possessions, since such players will not be associated with large changes in win probability. 
One way to overcome this bias is to account for which team has possession of the ball into our win probability estimates. 
In principle, it would be straightforward to include possession information into our win probability estimates: first we bin the games based on home team lead, time remaining, and which team has possession, and then we apply our estimation procedure twice, once for when the home team has possession and once for when the away team has possession.
Our omission of possession information is driven largely by our inability to determine which team has possession on a second-by-second basis reliably due to errors in the order in which plays are recorded in the play-by-play data we have used.
In general, more sophisticated estimation of win probability remains an area of future exploration.

Since our estimates of player effect are context-dependent, we have introduced leverage profiles as a way to determine which players' partial effects are most directly comparable.
Though we have not done so in this paper, one could use leverage profiles to cluster players based on the situations in which they play. 
This could potentially provide insight into how coaches use various players around the league and lead to a more nuanced understanding of each player's role on his team.

In keeping with the spirit of previous player-evaluation, we define two metrics, Impact Ranking and Impact Score, to determine a rank-ordering of players.
Impact Ranking provides an in-team ranking of each player's partial effect, allowing us to determine whether a player's salary is commensurate with his overall contribution to his team's chances of winning games.
Impact Score balances a player's estimated effect against the uncertainty in our estimate to generate a league-wide rank-ordering.

We have found that any individual player's effect on his team's chances of winning during a single shift is small, generally less than 1\%. 
We moreover have found rather considerable overlaps in the posterior distribution of player partial effects.
This suggests there is no single player who improves his team's chances of winning significantly more than the other players.
That said, we are still able to distinguish clear differences in players' impacts.
Somewhat surprisingly, we find that Dirk Nowitzki had a larger impact on his team's chances of winning that more prominent players like Kevin Durant and LeBron James.
We also found that Durant and James' impact were virtually indistinguishable.
This is not to suggest that Nowitzki is a better or more talented player than Durant or James, per se.
Rather, it indicates that Nowitzki's performance was much more important to his team's success than Durant's or James' performances were to their respective teams.

There are several possible extensions and refinements to our proposed methodology.
As mentioned earlier, our win probability estimation is admittedly simplistic and designing a more sophisticated procedure is an area for future work.
It is also possible to include additional covariates in equation~(\ref{eq:player_team_model}) to entertain two-way or three-way player interactions, in case there are any on-court synergies or mismatches amongst small groups of players.
In its current form, Equation~\ref{eq:player_team_model} does not distinguish the uncertainty in estimating the $y_{i}$'s from the inherent variability in the change in win probability.
It may be possible to separate these sources of variability by decomposing $\sigma,$ though care must be taken to ensure identifiability of the resulting model. 
Finally, rather than focusing on each player's overall impact, one could scale the predictors in Equation~\ref{eq:player_team_model} by the shift length and re-fit the model to estimate each player's per-minute impact on his team's chances of winning.

\newpage
\bibliography{basketball}

\begin{thebibliography}{}

\bibitem[Ballentine, 2014]{Ballentine2014}
Ballentine, A. (2014).
\newblock {K}evin {D}urant {C}ements {H}imself as {MVP} {F}rontrunner with
  {T}riple-{D}ouble in {R}eturn.

\bibitem[Bashuk, 2012]{Bashuk2012}
Bashuk, M. (2012).
\newblock {U}sing {C}umulative {W}in {P}robability to {P}redict {NCAA}
  {B}asketball {P}erformance.
\newblock In {\em Sloan Sports Analytics Conference}, pages 1--10.

\bibitem[Buckley, 2014]{Buckley2014}
Buckley, Z. (2014).
\newblock {L}e{B}ron {J}ames {R}eminding {E}veryone 2013 {MVP} {R}ace is {F}ar
  from {O}ver.

\bibitem[Cavan, 2014]{Cavan2014}
Cavan, J. (2014).
\newblock {A}re {NBA} {T}eams {O}vervaluing {S}tretch 4s in {F}ree {A}gency.

\bibitem[Goldsberry and Weiss, 2013]{GoldsberryWeiss2013}
Goldsberry, K. and Weiss, E. (2013).
\newblock {T}he {D}wight {E}ffect: {A} {N}ew {E}nsemble of {I}nterior {D}efense
  {A}nalytics for the {NBA}.
\newblock In {\em Sloan Sports Analytics Conference}, pages 1--11.

\bibitem[Gramacy et~al., 2013]{GramacyJensenTaddy2013}
Gramacy, R.~B., Jensen, S.~T., and Taddy, M. (2013).
\newblock {E}stimating {P}layer {C}ontribution in {H}ockey with {R}egularized
  {L}ogistic {R}egression.
\newblock {\em Journal of Quantitative Analysis in Sports}, 9:97--111.

\bibitem[Hollinger, 2004]{Hollinger2004}
Hollinger, J. (2004).
\newblock {\em {P}ro {B}asketball {F}orecast: 2004-05 {E}dition}.
\newblock Potomac Books.

\bibitem[Ilardi, 2014]{Ilardi2014}
Ilardi, S. (2014).
\newblock {T}he {N}ext {B}ig {T}hing: {R}eal {P}lus-{M}inus.

\bibitem[Ilardi and Barzilai, 2008]{IlardiBarzilai2008}
Ilardi, S. and Barzilai, A. (2008).
\newblock {A}djusted {P}lus-{M}inus {R}atings: {N}ew and {I}mproved for
  2007-2008.

\bibitem[Kyung et~al., 2010]{KyungGillGhoshCasella2010}
Kyung, M., Gill, J., Ghosh, M., and Casella, G. (2010).
\newblock {P}enalized {R}egression, {S}tandard {E}rrors, and {B}ayesian
  {L}assos.
\newblock {\em Bayesian Analysis}, 5(2):369--412.

\bibitem[Lapin, 2014]{Lapin2014}
Lapin, J. (2014).
\newblock {U}nder-the-{R}adar {F}ree-{A}gent {B}argains {H}ouston {R}ockets
  {M}ust {C}onsider.

\bibitem[Lindsey, 1963]{Lindsey1963}
Lindsey, G. (1963).
\newblock An {I}nvestigation of {S}trategies in {B}aseball.
\newblock {\em Operations Research}, 11(4):477--501.

\bibitem[Maymin et~al., 2012]{MayminMayminShen2012}
Maymin, A., Maymin, P., and Shen, E. (2012).
\newblock {H}ow {M}uch {T}rouble is {E}arly {F}oul {T}rouble? {S}trategically
  {I}dling {R}esources in the {NBA}.
\newblock {\em International Journal of Sport Finance}, 7:324--339.

\bibitem[Mills and Mills, 1970]{MillsMills1970}
Mills, E.~G. and Mills, H.~D. (1970).
\newblock {P}layer {W}in {A}verages: {A} {C}omplete {G}uide to {W}inning
  {B}aseball {P}layers.
\newblock In {\em The Harlan D. Mills Collection}.

\bibitem[Park and Casella, 2008]{ParkCasella2008}
Park, T. and Casella, G. (2008).
\newblock {T}he {B}ayesian {L}asso.
\newblock {\em Journal of the American Statistical Association},
  103(482):681--686.

\bibitem[Pettigrew, 2015]{Pettigrew2015}
Pettigrew, S. (2015).
\newblock {A}ssessing the {O}ffensive {P}roductivity of {NHL} {P}layers {U}sing
  {I}n-{G}ame {W}in {P}robabilities.
\newblock In {\em Sloan Sports Analytics Conference}, pages 1--9.

\bibitem[Rosenbaum, 2004]{Rosenbaum2004}
Rosenbaum, D. (2004).
\newblock {M}easuring {H}ow {NBA} {P}layers {H}elp {T}heir {T}eams {W}in.

\bibitem[Stern, 1994]{Stern1994}
Stern, H. (1994).
\newblock A {B}rownian {M}otion {M}odel for the {P}rogress of {S}ports
  {S}cores.
\newblock {\em Journal of the American Statistical Association}, 89(427):1128
  -- 1134.

\bibitem[Studeman, 2004]{Studeman2004}
Studeman, D. (2004).
\newblock {T}he {O}ne {A}bout {W}in {P}robability.

\bibitem[Thomas et~al., 2013]{ThomasVenturaJensenMa2013}
Thomas, A., Ventura, S.~L., Jensen, S.~T., and Ma, S. (2013).
\newblock {C}ompeting {P}rocess {H}azard {F}unction {M}odels for {P}layer
  {R}atings in {I}ce {H}ockey.
\newblock {\em The Annals of Applied Statistics}, 7(3):1497 -- 1524.

\bibitem[Tibshirani, 1996]{Tibshirani1996}
Tibshirani, R. (1996).
\newblock {R}egression {S}hrinkage and {S}election via the {L}asso.
\newblock {\em Journal of the Royal Statistical Society. Series B
  (Methodological)}, 85(1):267--288.

\bibitem[Zou and Hastie, 2005]{ZouHastie2005}
Zou, H. and Hastie, T. (2005).
\newblock {R}egularization and {V}ariable {S}election via the {E}lastic {N}et.
\newblock {\em Journal of the Royal Statistical Society. Series B
  (Methodological)}, 67(301-320).

\end{thebibliography}

\newpage
\section*{Appendix}

\begin{figure}[!h]
\centering
\includegraphics{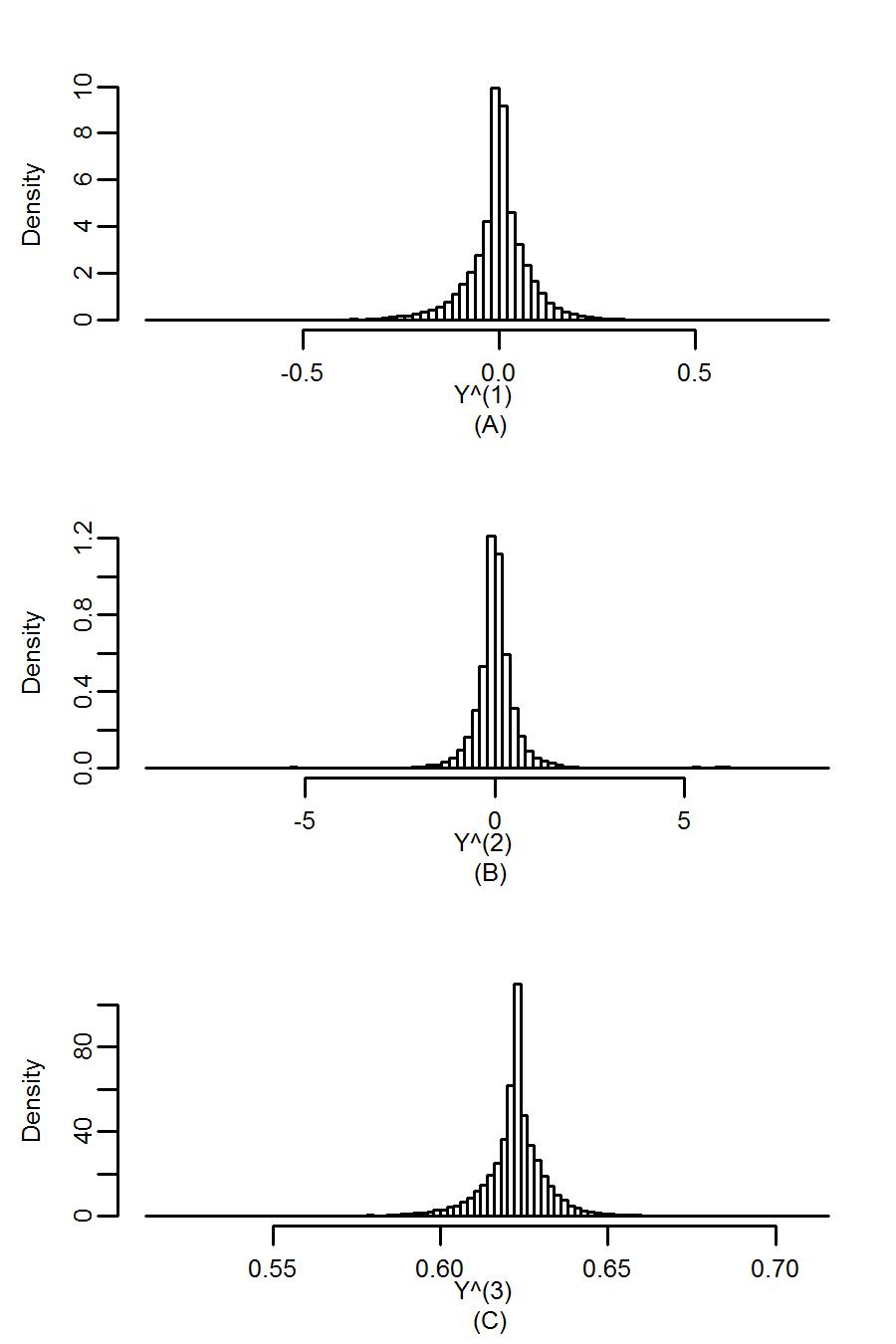}
\caption{Histogram of $y^{(1)}$ (A),  $y^{(2)}$  (B), and $y^{(3)}$(C).}
\label{fig:transform_histograms}
\end{figure}

As we discussed in Section~\ref{sec:data_model_methods}, we have made several strong assumptions in specifying a Gaussian linear regression model.
We now check and discuss the assumption that the errors in Equation~\ref{eq:player_team_model} are Gaussian with constant variance.
We also consider several transformations and alternative model specifications which could potentially align with these assumptions better than our original specification.
In particular, we consider the following response variables, $y^{(1)}, y^{(2)},$ and $y^{(3)}$:
\begin{itemize}
\item{$y_{i}^{(1)}$: our original response, the change in the win probability.}
\item{$y_{i}^{(2)}$: the change in the log-odds of winning the game. Intuitively, this further down-weights the importance of low-leverage performance as a 5\% change in win probability from 45\% to 50\% corresponds to a much larger change in the log-odds than a 5\% change in win probability from 90\% to 95\%.}
\item{$y_{i}^{(3)} = \left[ 1 + \exp{\left\{-\frac{1 + y_{i}}{2}\right\}}\right]^{-1}$: the inverse logit transformation of the shifted and re-scaled change in win probability.}
\end{itemize}
Figure~\ref{fig:transform_histograms} shows histograms of these response variables, along with a histogram of our original response, change in win probability.

\begin{figure}[!h]
\centering
\includegraphics{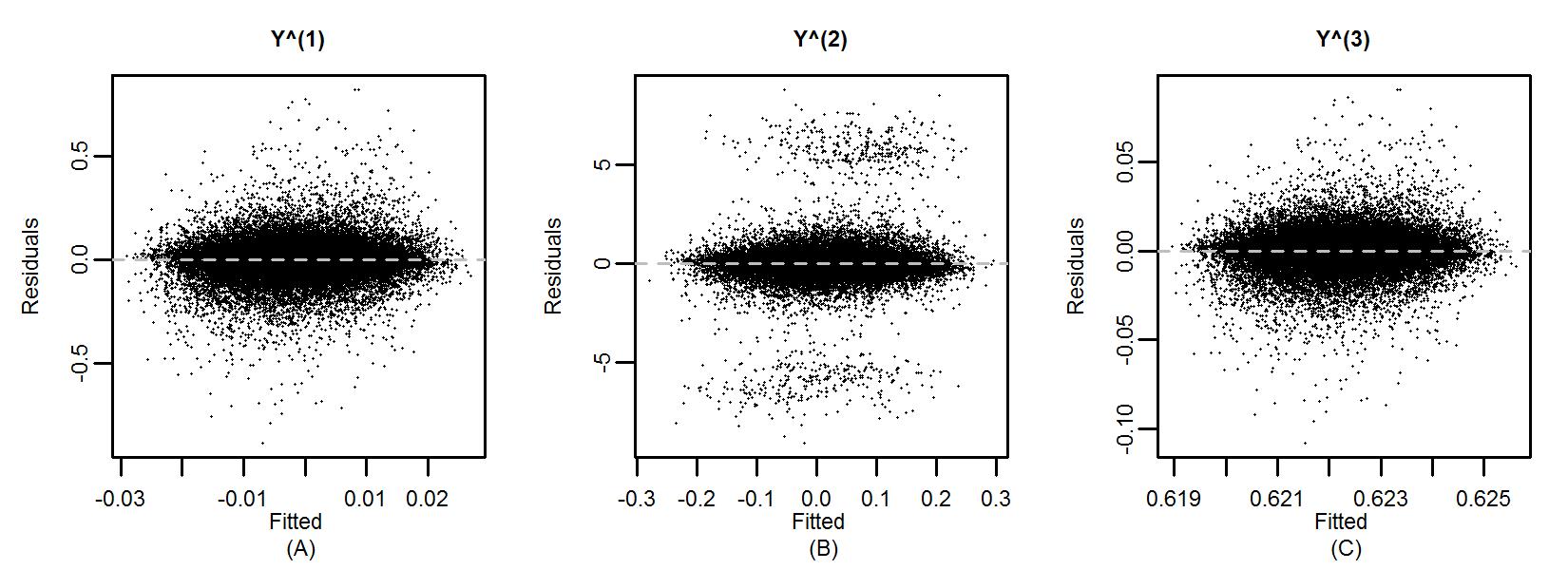}
\caption{Residuals plotted against fitted variable for the original model, the log-odds model, and the inverse logit model}
\label{fig:transform_residuals}
\end{figure}

We notice that the distribution of $y^{(1)}$ and the distribution of $y^{(3)}$ are similar: both are rather tightly concentrated near 0 and 0.622, respectively and are more or less symmetric.
We also observe that $y^{(3)}$ is much less variable than $y^{(1)}.$ 
Interestingly, we see in Figure~\ref{fig:transform_histograms}(B), that the change in the log-odds of winning is slightly more heavy-tailed than these other distributions. 
In particular, we see that in about 2\% of all shifts, the absolute value of the change in the log-odds of winning exceeds 5. 
These correspond to shifts in which there was a very large swing in the home team's win probability during a given shift.
For example, in the penultimate shift of the March 16, 2014 game between the Miami Heat and the Houston Rockets, the Heat went from trailing by 5 points with 6:13 left to leading by 9 points with a few seconds left.
In doing so, the Heat increased their win probability from 22\% to 98\%, corresponding to a change in the log-odds of winning of about 5.86.
Given the fact that for the vast majority of shifts that the change in win probability and the change in the log-odds of winning the game were very close to 0 (indicated by the large ``spikes'' in the histograms near 0) and the fact that we have imposed rather strong shrinkage on our player effects, we would not expect our model to be able to estimate such a large change in the log-odds reliably. 
This is borne out in the residual plot, show in Figure~\ref{fig:transform_residuals}(B): these shifts had residuals near $\pm 5.$
To find the fitted values in Figure~\ref{fig:transform_residuals}, we first simulated 1000 posterior draws of the conditional expectation function $E[y^{(\cdot)}_{i}|\mathbf{P}_{i}, \mathbf{T}_{i}] = \mathbf{P}_{i}^{\top}\theta + \mathbf{T}_{i}^{\top}\tau$ using our simulated posterior draws of $\theta$ and $\tau$ and then took the average.
It is important to note that because of the regularization, the fitted values and residuals are biased so we do not expect that they will be Gaussian.

\begin{figure}[!h]
\centering
\includegraphics{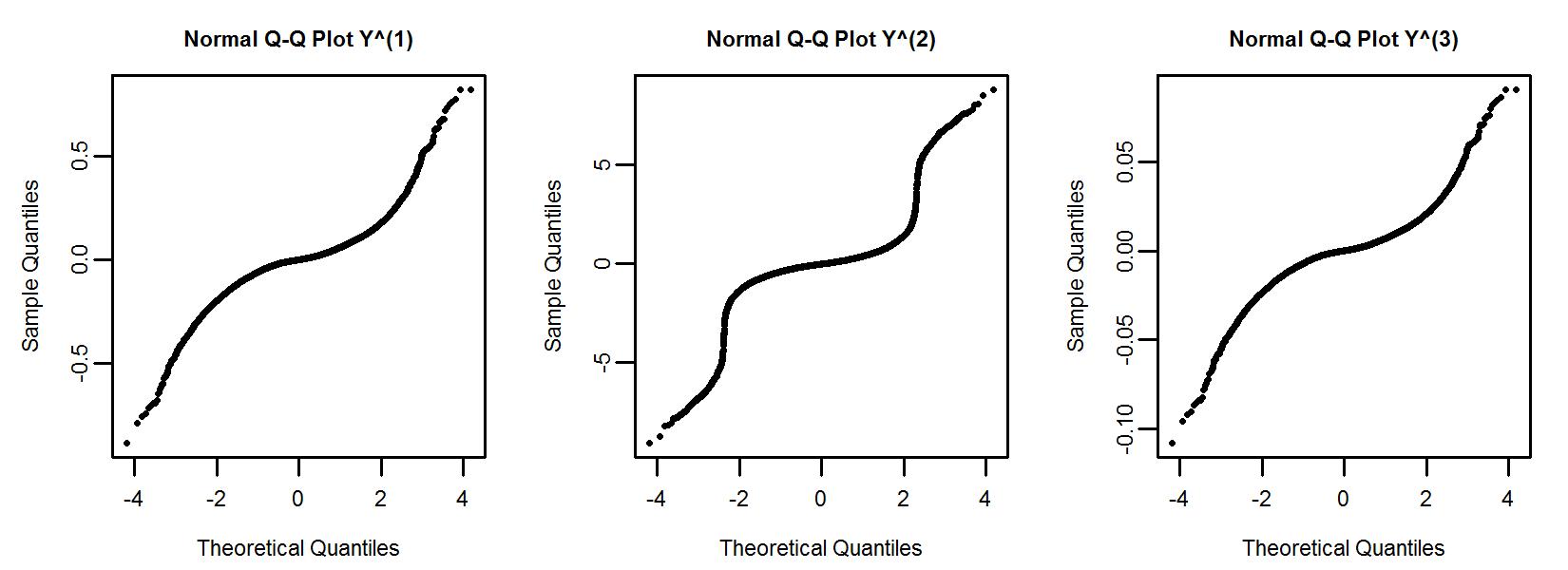}
\caption{Normal quantile plots of residuals from modeling $y^{(1)}$ (A), $y^{(2)}$ (B), $y^{(3)}$ (C)}
\label{fig:qq_norm_transform_residuals}
\end{figure}

We note that Figures~\ref{fig:transform_residuals}(A) and (C) are very similar in shape, though we note that the residuals in (C) are much smaller.
This is not particularly surprising, since the variance of $y^{(3)}$ is much smaller than the variance of $y^{(1)}.$
Just like we might in a standard least squares regression problem, we can form normal quantile plots of these residuals. 
It is important to note, however, that the residuals are not unbiased estimators of the error terms because of the regularization.
Nevertheless, it may still be desirable to consider an alternative model specification in which the distribution of the resulting residuals is much closer to Gaussian than our original model. 
Figure~\ref{fig:qq_norm_transform_residuals} shows the resulting normal quantile plots.
As anticipated, we see that none of the residual plots display the linear trend characteristic of Gaussian distributions. 
Interestingly, we also observe that the residuals corresponding to $y^{(2)}$ seem decidedly less Gaussian and the residuals corresponding to $y^{(3)}$ appear to be similar to our original residuals. 
In light of this, we do not find the suggested transformations particularly compelling, in terms of aligning with our original modeling assumptions.

\begin{figure}[!h]
\centering
\includegraphics[width = 4.5in]{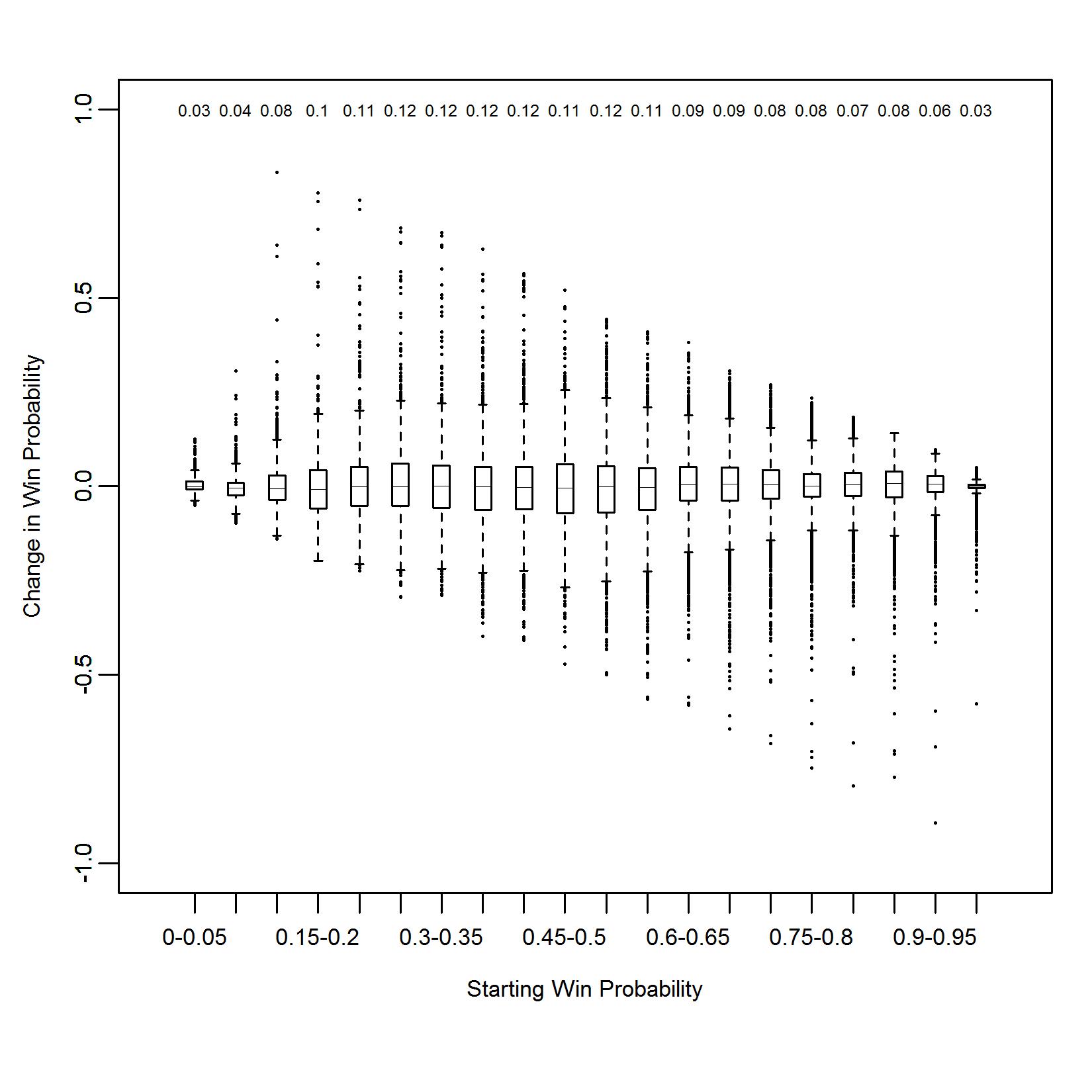}
\caption{Change in win probability binned according to the starting win probability of the shift. Since win probability is constrained to the interval [0,1], as the win probability at start of the shift increases, the distribution of the change in win probability shifts from right-skewed to left-skewed.}
\label{fig:y_bin_sd}
\end{figure}

We now consider the issue of homoscedasticity.
Once again, we note that we are assuming that, conditional on the players on the court, the variance of the change in win probability is constant.
Since we only observe a handful of observations with the same ten players on the court, we cannot check this assumption prior to fitting our model.
Still, it is reasonable to suspect that the variance of $y_{i}$ depends on the win probability at the start of the shift.
Figure~\ref{fig:y_bin_sd} shows box plots of the change in win probability binned according to the starting win probability of the shift, along with the standard deviation of the observations in each binned.

We see immediately that the binned standard deviations are not constant, indicating that the variance of $y^{(1)}$ does depend on the starting win probability.
However, we see that this dependence really only manifests itself when the starting win probability is close to 0 or 1.
It is worth mentioning that this dependence in and of itself does not invalidate our initial assumption that the variance of $y^{(1)}$ conditional on the players on the court is constant. 
However, it does suggest that we try re-weighting the response and predictors in Equation~\ref{eq:player_team_model} so that the binned standard deviations of the re-weighted response are constant.
This is similar to what we might do in a weighted least squares regression problem. 
We consider three re-weighting schemes:
\begin{itemize}
\item{Re-scale so that the binned standard deviations are all 1. This magnifies the response and predictors for all shifts. Denote the new response variable $y^{(4)}.$}
\item{Re-scale so that the binned standard deviations are all 0.03. This shrinks the response and predictors for all high-leverage shifts but leaves the observations from low-leverage shifts relatively unchanged. Denote the new response variable $y^{(5)}$.}
\item{Re-scale so that the binned standard deviations are all 0.12. This magnifies the response and predictors for all low-leverage shifts but leaves the observations from high-leverage shifts relatively unchanged. Denote the re-scaled response variable $y^{(6)}.$}
\end{itemize}

\begin{figure}[!h]
\centering
\includegraphics{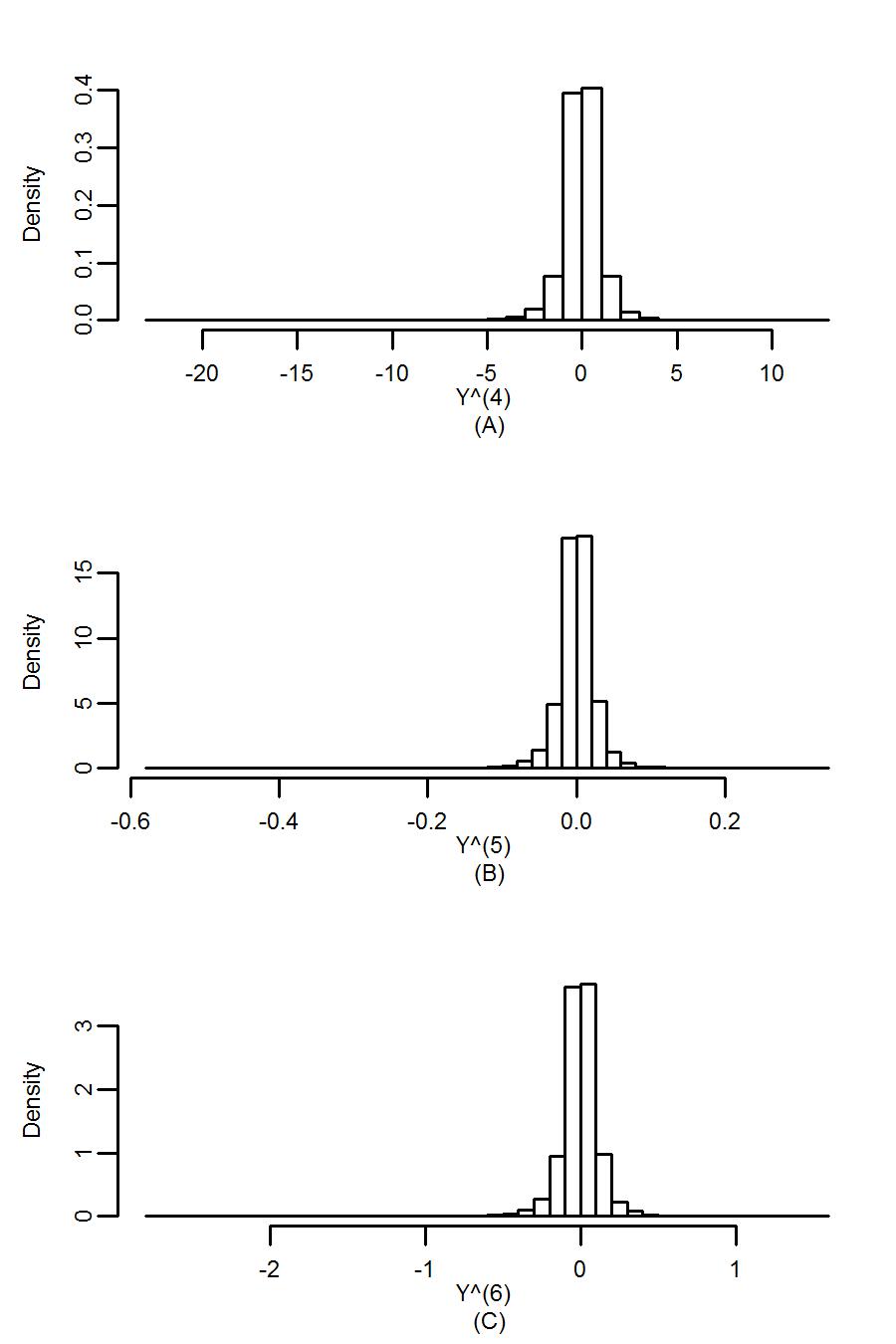}
\caption{Histograms for $y^{(4)}, y^{(5)},$ and $y^{(6)}.$}
\label{fig:re_scaled_histograms}
\end{figure}

\begin{figure}[!h]
\centering
\includegraphics{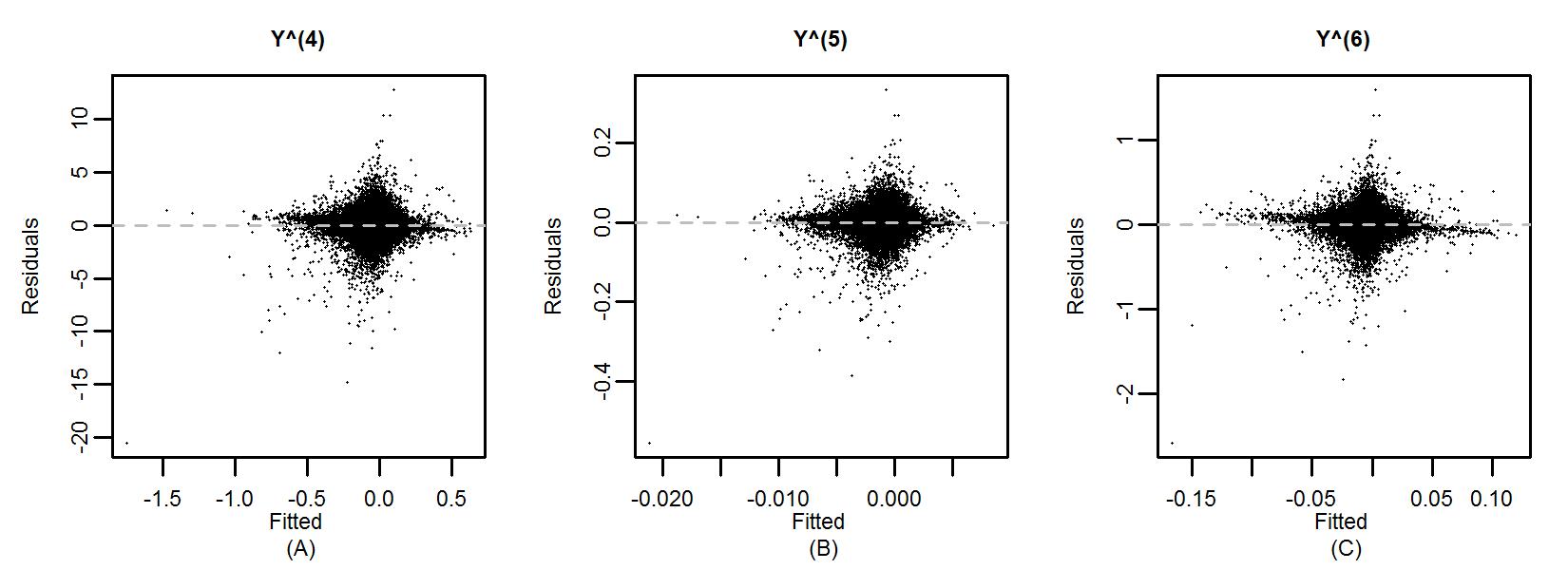}
\caption{Residuals resulting from modeling $y^{(4)}, y^{(5)}, y^{(6)}$}
\label{fig:re_scaled_residuals}
\end{figure}

Figure~\ref{fig:re_scaled_histograms} shows histograms of the three re-scaled responses and Figure~\ref{fig:re_scaled_residuals} show the residuals that arise from fitting the three re-scaled models.
Though it is not immediately apparent in Figure~\ref{fig:re_scaled_residuals}, we observe rather extreme values of $y^{(4)}$ like -22.62, mainly corresponding to late-game shifts during which the win probability changed dramatically. 
We find that $y^{(5)}$ and $y^{(6)}$ are somewhat more tightly concentrated near 0 than is $y^{(1)}.$

Figure~\ref{fig:re_scaled_residuals} plots the residuals against the fitted values from the re-scaled models. 
Once again, these residuals are biased, so we do not expect them to resemble the residuals plots obtained in standard least squares regression problems. 
The noticeable negative trend, seen especially in Figures~\ref{fig:re_scaled_residuals}(A) and (C), is a good indication of the bias introduced by regularization.
It is interesting that this bias is much more apparent after re-scaling the response and predictors than it was in our original model.

Somewhat worryingly, we see that there is considerable variation in the residuals when the fitted value is near zero. 
This is in marked contrast to Figure~\ref{fig:transform_residuals}(A), in which we see more or less constant variation in the residuals for all possible fitted values.
It appears that correcting for potential heteroscedasticity results in residuals that are even less well-behaved than in our original model.
As a result, we do not see any of the re-weighted models as being necessarily better than our original model.

\end{document}